\documentclass{article}
\pdfoutput=1

\usepackage{arxiv}

\usepackage{graphicx}
\usepackage{dcolumn}
\usepackage{amsmath,amsfonts,amssymb}  
\usepackage{bm}
\usepackage{color}
\usepackage{hyperref}
\usepackage{overpic}
\usepackage{colortbl,xcolor}
\usepackage{booktabs}
\usepackage{tikz}
\usepackage{pict2e}
\definecolor{green}{cmyk}{0.75002 0 1 0}
\definecolor{yellow}{cmyk}{0.04 0.3 1 0.02}
\definecolor{orange}{cmyk}{0 0.6 1 0}
\definecolor{blue}{rgb}{ 0    0.2    1.0}
\definecolor{red}{rgb}{1 0 0}

\title{Vertical vortex gust encounters of a delta wing}

\author{
  Zhecheng Liu\thanks{Email for correspondence: \texttt{zliu163@ucla.edu}}, Kunihiko Taira, and Jeff D. Eldredge\\
  Department of Mechanical and Aerospace Engineering\\
  University of California, Los Angeles, California 90095, USA\\
}

\date{\today}

\begin{document}

\maketitle

\begin{abstract}
The present study analyzes encounters of a vertical vortex with a tailless NACA 0012 delta wing with sweep angle $60^\circ$ at an angle of attack of $10^\circ$ and Reynolds number 1000 using direct numerical simulations. Motivated by the need to fly in complex settings such as urban canyons, mountainous areas, wildfires, or vehicle swarms, we focus on the influence of a vertical vortex gust on the delta wing, complementing previous efforts that predominantly studied spanwise and streamwise vortex encounters. The vertical orientation can impose strong transient loads on all six degrees of freedom, and particular attention is devoted to connecting these loads to the flow response. By varying the gust's lateral position, size, and strength, we find a fundamental process that characterizes vertical vortex gust encounters. The gust consistently imprints a low-pressure core on the wing surface. The evolving core creates an adverse pressure gradient that distorts the baseline skin-friction pattern and induces separation that lifts the boundary layers. These lifted boundary layers are elongated and twisted by the gust's strain into lobed vortical structures, which are shown by force element analysis to make a critical contribution to the lift. As the gust's lateral position is varied from root to tip, the low-pressure core and the induced flow separation are increasingly localized around the leading edge closest to the gust, modifying the lobe evolution and leading to complex dependence of the force and moment response on gust position compared to the largely monotonic dependence on gust size and strength.
\end{abstract}


\section{Introduction}
As small-scale air vehicles are increasingly deployed in surveillance and transportation, they are required to maintain a stable flight in complex and often extreme flight environments where the relative scales of the gusts can be disproportionally large, due to the small size and low cruise speed of these vehicles \cite{Jones_2022,taira2025extreme}. Such gust encounters can induce large transient forces and moments, potentially compromising the vehicle stability. Advancing our understanding of the underlying flow physics in gust–wing interactions is therefore essential to enable robust operation and control of small-scale air vehicles in complex environments.

Although many investigations of gust interactions in previous works have focused on a two-dimensional flow configuration, they have offered us valuable insights. For example, K\"{u}ssner \cite{Kussner1932} exploited the linearity of a small-amplitude gust encounter to derive a linear analytical model to predict the lift and moment changes when a two-dimensional flat plate enters a sharp-edged transverse gust. In their experimental study of a nominally two-dimensional flat plate traveling through a transverse jet, Sedky et al. \cite{Sedky_2022} superposed the K\"{u}ssner model and the Wagner model \cite{Wagner1925} to derive a linear controller that mitigated the excursion from zero lift.

However, a gust in a realistic flight environment usually comes in a more complex form (e.g., a series of vortical gusts), and it is essential to account for that complexity in large-amplitude encounters where linear models are relatively unreliable. Thus, complex gust encounters have stimulated some recent works, particularly with the objective of learning a reduced-order model from data. To date, these studies have mostly relied on two-dimensional airfoil encounters with discrete vortices to enable efficient training sets. Fukami and Taira \cite{Fukami2023} studied an encounter with a strong Taylor vortex, which induced massive flow separation at both the leading edge and trailing edge of a two-dimensional NACA 0012 airfoil. They proposed to reduce the flow and force responses via a lift-augmented autoencoder and found that the responses lay in a three-dimensional latent space during gust-airfoil encounters. Lopez-Doriga et al. \cite{lopezdoriga2025} explored the effect of the airfoil geometry on the Taylor vortex gust encounters. Building upon the autoencoder work, Liu et al. \cite{Liu2025} proposed an autoencoder-based framework of model-based reinforcement learning to find an effective control strategy for lift mitigation of a flat plate during discrete vortex gust encounters. Liu and Eldredge \cite{liu2025attention} found that, in lieu of using the full flow field, data from a small number of surface pressure sensors---reduced by a transformer---were sufficient for a reinforcement learning agent to successfully learn a generalizable control strategy for regulating lift, and that this strategy could be extended from a short sequence of gusts to an arbitrarily long sequence. Therefore, there are a significant amount of studies that investigated the complex vortical gust encounters in a two-dimensional configuration.

There have also been investigations of gust-wing interaction beyond a purely two-dimensional flow configuration. For example, Barnes and Visbal \cite{Barnes2018,Barnes2018_2} investigated how a spanwise Taylor vortex tube influences the boundary layer of a NACA 0012 airfoil during encounters at a transitional Reynolds number in a spanwise periodic flow setting. Fukami et al.~\cite{Fukami2025} also investigated the spanwise periodic setting with a spanwise Taylor vortex tube encounter by the NACA 0012 airfoil; they showed that the autoencoder methodology developed for a Reynolds number $100$ two-dimensional flow in their previous work \cite{Fukami2023} could be extended to a Reynolds number $5000$, spanwise periodic flow configuration. Odaka et al.~\cite{odaka_square_2026} explored a spanwise Taylor vortex tube encounter of both a two-dimensional airfoil and a three-dimensional square wing at Reynolds number $600$, and showed that the tip vortices of the square wing reduce the transient lift variations compared to the two-dimensional airfoil. Further work by Odaka et al.~\cite{Odaka2026} also revealed that the large-scale vortical structures during the vortex gust encounters of a square wing bear a striking similarity between $Re=600$ and $10000$, when the larger Reynolds number flow's structures are clarified by a scale decomposition.

These previous works primarily focused on spanwise vortex gust encounters, which induce relatively uniform disturbance along the wing span, and because of symmetry, have no effect on the side force and the associated roll and yaw moment. A few studies have investigated streamwise vortex gust encounters of a rectangular flat plate wing \cite{Garmann2015,Barnes2015,McKenna2017}. For example, Garmann and Visbal \cite{Garmann2015} characterized the unsteady modes of the interactions resulting from different spanwise positions of the incoming streamwise vortex. However, gusts in realistic flow environments, such as strong atmospheric turbulence, wakes of urban buildings, above wildfires, or in vehicle swarms, are inherently three-dimensional and exhibit a wide range of orientations \cite{Hu_Morgans_2022,Hu_Morgans_2023,Adrian_2007,Fernando_2019,Ali2018}. In the particular case of a small air vehicle flying in the wake of a high-rise urban building, it is expected that vortical structures will be predominantly vertically oriented \cite{Hu_Morgans_2022,Hu_Morgans_2023}. Since all previous works have focused on spanwise or streamwise vortex gust encounters, a study of vertical vortex gust encounters is essential and urgently needed.

This study focuses on a vertical vortex encounter with a tailless delta wing. Delta wing aerodynamics are particularly sensitive to external disturbances because their highly-swept geometry gives rise to a vortex-dominated lift mechanism. Because of this sweep, the wing's aerodynamic performance is closely tied to the stability of its leading-edge vortices (LEVs), so disturbances that modify these coherent structures can induce pronounced changes in the aerodynamic loads and thereby challenge the flight stability and control. A significant number of previous studies have investigated the aerodynamic performance and the associated vortical structures for different delta wing geometries and flow regimes in the absence of gusts \cite{Gursul2005_1,Gursul2005_2,Ol2003,Taylor2004,LeProvost2018}. The flight stability challenge becomes even more critical for tailless delta wings, since the absence of a tail reduces the passive stability and control authority available to counteract the coupled roll, yaw, and pitch responses induced by aerodynamic disturbances. A few experimental studies have investigated the gust effect on a tailless delta wing \cite{He2023,Chen2026,Marzanek2019}. For example, He and Williams \cite{He2023} experimentally revealed that a transverse gust induces substantial aerodynamic loads on the delta wing, especially in the form of roll moments associated with the asymmetric surface pressure distribution. However, the gusts in these studies were restricted to either streamwise or transverse irrotational velocity disturbances, and might not be sufficiently descriptive of gust encounters in realistic flight environments.

To address the aforementioned gap, we perform direct numerical simulations (DNS) to study the interactions between a NACA $0012$ tailless delta wing and a isolated straight, vertically-oriented Taylor vortex tube at Reynolds number $1000$ based on the root chord length. The delta wing has a sweep angle $\Lambda = 60^\circ$ and an angle of attack $\alpha=10^\circ$. Note that the geometry of the present delta wing is motivated by the D90 test vehicle, designed by Williams et al.~\cite{Williams2025,Williams2025_2} for exploration of the flight control in modern unmanned air vehicles. We systematically conduct a parametric study by varying the gust lateral position, size, and strength. By analyzing the imprinted surface signatures, flow kinematics, and force element decomposition, we illuminate the fundamental physical mechanism governing vertical vortex gust encounters. We further reveal how each individual gust parameter influences the interaction and the associated aerodynamic response. The rest of the paper is structured as follows. The problem set-up and verification study are provided in section \ref{sec:setup}. The results and discussions are given in section \ref{sec:parametric}, which starts by examining the fundamental difference of the vertical vortex gust encounters from a spanwise vortex gust encounters in section \ref{subsec:phi}. Section \ref{subsec:z0} reveals the physical mechanism that governs the vertical vortex gust encounters from three perspectives---imprinted surface signatures and flow kinematics (section \ref{subsec:surface_signatures}) and force element analysis (section \ref{subsec:force_element})---by inspecting gust encounters at three representative lateral positions. The effect of the gust size and strength is given in section \ref{subsec:DandG}. Finally, the conclusions are provided in section \ref{sec:conclusions}.


\section{Problem set-up}\label{sec:setup}
\subsection{Simulation set-up}
\begin{figure}[htbp]
    \centering
    \begin{overpic}[width=1.0\linewidth]{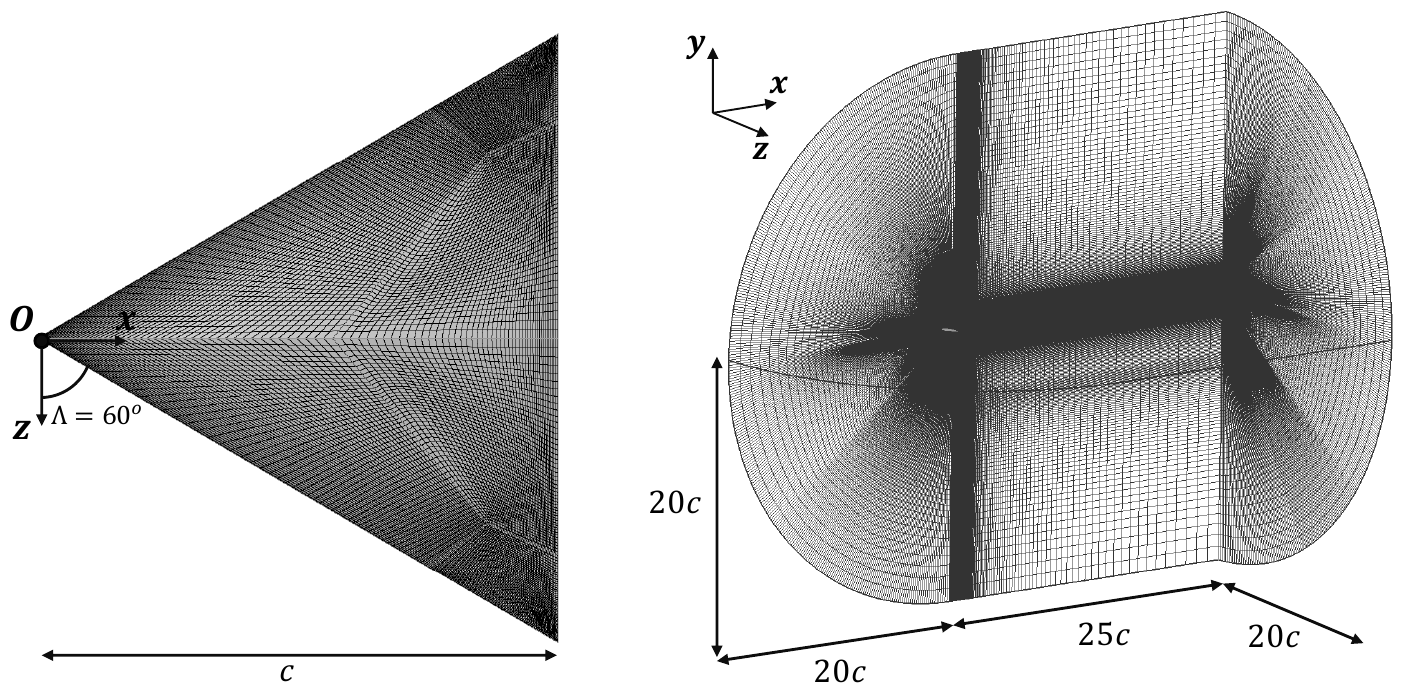}
    \put(0,47){{$(a)$}}
    \put(45,47){{$(b)$}}
    \end{overpic}
    \caption{$(a)$ the top view of the delta wing and the surface mesh; $(b)$ a perspective view of the computational domain.}
    \label{fig:mesh}
\end{figure}

We perform DNS of flows past a NACA $0012$ tailless delta wing under gust encounters. The present delta wing has a sweep angle $\Lambda = 60^\circ$ and an angle of attack $\alpha=10^\circ$ as shown in figure \ref{fig:mesh}. The three-dimensional flows are studied by numerically solving the incompressible Navier-Stokes equations:
\begin{gather}
    \frac{\partial \boldsymbol u}{\partial t} + \boldsymbol u \cdot \boldsymbol \nabla \boldsymbol u = - \boldsymbol \nabla p + \frac{1}{Re} \boldsymbol \nabla^2 \boldsymbol u, \\
    \boldsymbol \nabla \cdot \boldsymbol u = 0,
\end{gather}
where $\boldsymbol u = (u_x, u_y, u_z)$ is the velocity vector and $p$ is the pressure. We nondimensionalize the spatial variables by the root chord length $c$ of the delta wing, velocities by the freestream $U_\infty$, time by $c/U_\infty$, and pressure by $\rho U_\infty^2$, where $\rho$ is density. An incompressible flow solver \textit{Cliff} (in \textit{CharLES} software package, Cascade Technologies, Inc.), based on the finite-volume method with second-order accuracy in both time and space \cite{Ham2004,Ham2006}, is used throughout the present study.

We set the Reynolds number to $Re \equiv U_\infty c/\nu=1000$, where $\nu$ is the kinematic viscosity, to ensure the flow remains laminar. We note that the mean-chord-based Reynolds number in this configuration is then $Re_{\bar{c}} \equiv U_\infty \bar{c} / \nu = 500$, where $\bar{c}$ is the chord length at the spanwise location $z=b/2$ with $b$ defined as the half-span length. The aerodynamic-mean-chord Reynolds number is $Re_{\text{mac}} \equiv U_\infty c_{\text{mac}}/\nu=667$, where the mean aerodynamic chord is defined as $c_{\text{mac}}\equiv \frac{2}{A_w} \int_0^bc(z)dz$ with the planform area $A_w= \frac{\sqrt{3}}{3}c^2$ and $c(z)$ defined as local chord length at spanwise coordinate $z$.  The coefficients of drag $C_D$, lift $C_L$, and side $C_S$ components of the aerodynamic force are normalized by $\frac{1}{2}\rho U_\infty^2 A_w$. Similarly, the aerodynamic moment coefficients (i.e., roll moment $C_{M_r}$, yaw moment $C_{M_y}$, and pitch moment $C_{M_p}$) are normalized by $\frac{1}{2}\rho U_\infty^2 A_w c$. The reference point of the aerodynamic moment coefficients is set as the quarter-chord point in the root section, with coordinates $\boldsymbol x_{ref} = (0.25c \cos \alpha, -0.25c \sin \alpha, 0)$. As figure \ref{fig:mesh} shows, the computational domain covers $(x,y,z)=[-20,25] \times [-20,20] \times [-20,20]$ with the Cartesian coordinate origin positioned at the apex of the delta wing. The domain choice is sufficiently large to ensure a very low blockage ratio. A C-type grid topology is chosen with refinement in the vicinity of the wing and the wake region.

The simulation starts from a uniform freestream $\boldsymbol u=(U_\infty,0,0)$, and the inlet and farfield boundary conditions are also prescribed with the freestream. A convective boundary condition $\partial \boldsymbol u / \partial t + U_\infty \partial \boldsymbol u / \partial x = \boldsymbol 0$ is used at the outlet to allow wake structures to leave the domain without disturbing the near-field flow. A no-slip boundary condition is used for the wing surface. We run the simulation for $38.7$ convective time units to allow any initial transients to pass before the introduction of a gust. We note that the local Courant–Friedrichs–Lewy (CFL) number is restricted to be less than $1$ throughout the study, which is defined as $\text{CFL}=\frac{\Delta t}{2V_{\text{cell}}}\sum_f |\boldsymbol u_f \cdot \boldsymbol n_f|A_f$ with $\Delta t$ the time step size, $V_{\text{cell}}$ the cell volume, $\boldsymbol{u}_f$ the fluid velocity, $\boldsymbol{n}_f$ the normal vector, and $A_f$ the surface area of the face $f$. The details regarding our choice of grid resolution, along with the convergence studies, are provided in the appendix \ref{app:appendix}.

\subsection{Gust implementation}
\begin{figure}[htbp]
    \centering
    \begin{overpic}[width=1.0\linewidth]{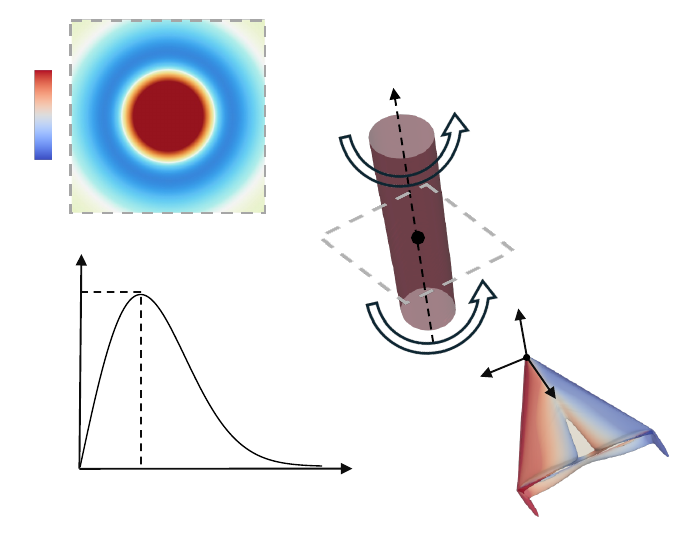}
    \put(19.5,6.7){\large{$R$}}
    \put(51,10){\Large{$r$}}
    \put(7.5,40){\Large{$u_{\theta}$}}
    \put(5,35){\Large{$u_{\theta_{\max}}$}}
    \put(1,60){\Large{$\omega_v$}}
    \put(4.2,52){\large{$-3$}}
    \put(5.6,68){\large{$3$}}
    \put(81.5,20){\Large{$x$}}
    \put(76.9,32){\Large{$y$}}
    \put(69,24.5){\Large{$z$}}
    \put(77.4,25.4){\large{$O$}}
    \put(59,64.5){\Large{$\boldsymbol{e}_v$}}
    \put(66.5,42.5){\Large{$\boldsymbol{x}_0=(x_0,y_0,z_0)$}}
    \end{overpic}
    \caption{The instantaneous $Q$ isosurface coloured by the vortex-axis direction vorticity component $\omega_v$, the corresponding contours of $\omega_v$ in an $\boldsymbol{e}_v$-normal slice, and the angular velocity profile of the Taylor vortex.}
    \label{fig:gust}
\end{figure}
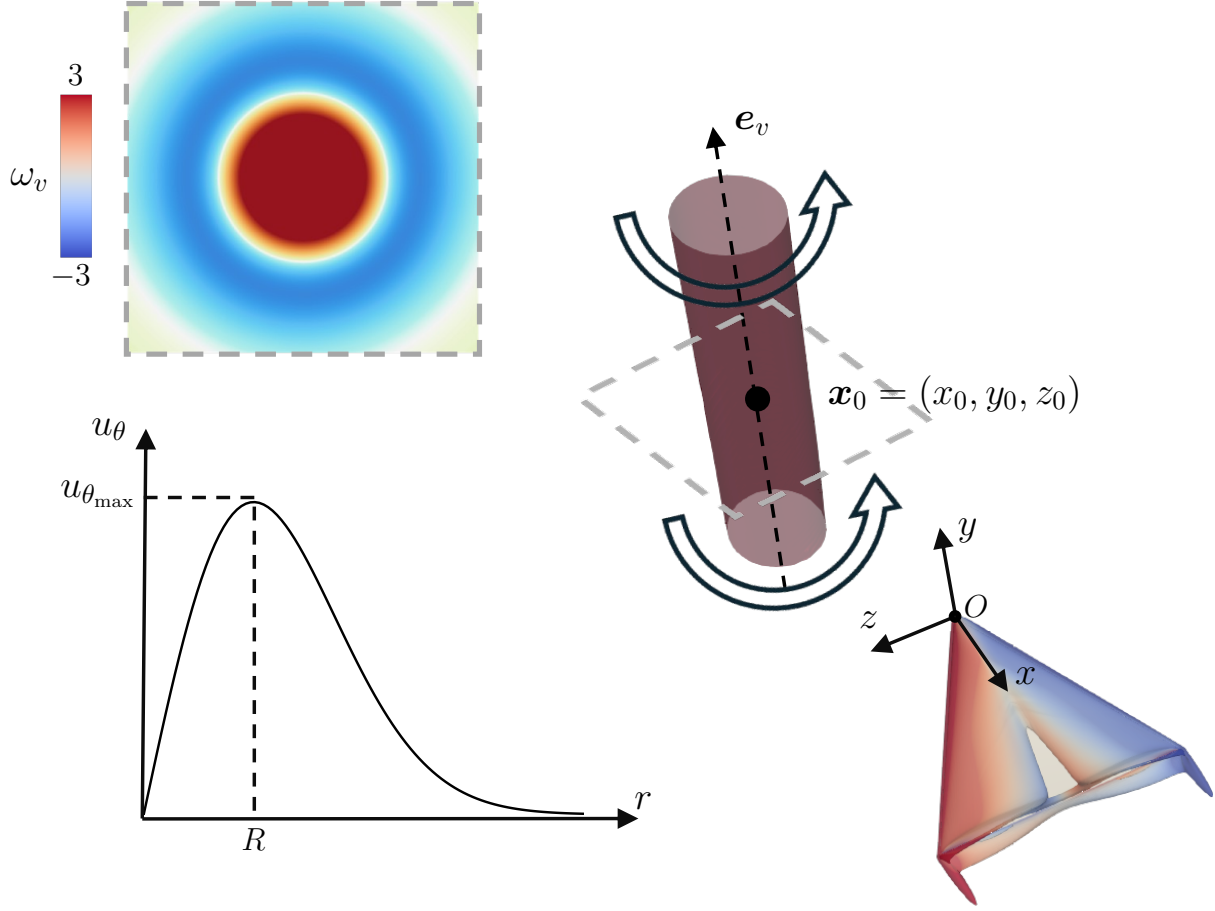

Each vortex gust is introduced upstream of the wing in the form of a Taylor vortex, as shown in figure \ref{fig:gust}. The Taylor vortex has an angular velocity profile given by
\begin{equation}\label{eq:taylorvortex}
    u_\theta = u_{\theta_{\max}}\frac{r}{R}\exp \left[ \frac{1}{2} \left( 1-\frac{r^2}{R^2} \right)\right],
\end{equation}
where a maximum angular velocity $u_{\theta_{\max}}$ is reached at radial distance $r=R$ from the vortex axis, where $R$ is the radius of the Taylor vortex. We can thus define the strength and the size of the Taylor vortex by prescribing $u_{\theta_{\max}}$ and $R$, respectively. In the three-dimensional configuration, a vortical tube is constructed with equation \eqref{eq:taylorvortex}, as shown in figure \ref{fig:gust}. We define the tube-axis direction $\boldsymbol{e}_v$ restricted within the $y-z$ plane, where its angle to the positive $z$ axis is defined as the gust roll angle $\phi$ (measured about the negative $x$ axis, as conventional for roll). For clarity, $\phi=0$ and $\pi$ both represent spanwise gust vortex tubes aligned with the $z$ axis, inducing initial upwash or downwash at the wing's apex, respectively. In contrast, $\phi=\pi/2$ and $3\pi/2$ both represent vertical gust vortex tubes aligned with the $y$ axis, inducing flow that is counter-clockwise or clockwise, respectively, when viewed from above. A reference point on the vortex axis is described by $\boldsymbol x_0 = (x_0,y_0,z_0)$. By prescribing the axis direction and the axis reference point, we fully specify the initial orientation and position of the vortex.

In the present study, the gust effect on the flow response is examined by sweeping over the parameters of gust ratio $G \equiv u_{\theta_{\max}} \in [0.1,1.5]$, the size $D \equiv 2R \in [0.25,1.25]$, the roll orientation $\phi \in [0, 2\pi)$, and the lateral position $z_0 \in [0,0.6]$ (where all parameters have been nondimensionalized as described above). We set $x_0=-1.5$ and $y_0=0.1$ throughout the study. The set of parameters considered in the present study is summarized in table \ref{tab:gust_parameters}.

\begin{table}[htbp]
\centering
\caption{The parameter set of the gust in the present study.}
\label{tab:gust_parameters}
\begin{tabular}{cccccccccccccccccccc}
\toprule
Case ID & $G$ & $D$ & $\phi$ & $z_0$ &
Case ID & $G$ & $D$ & $\phi$ & $z_0$ &
Case ID & $G$ & $D$ & $\phi$ & $z_0$\\
\midrule
$1$ & $1$ & $0.5$ & $0$ & $0$ & $9$ & $1$ & $0.5$ & $3\pi/2$ & $0.15$ & $17$ & $0.1$ & $0.5$ & $\pi/2$ & $0$ \\
$2$ & $1$ & $0.5$ & $\pi/2$ & $0$ & $10$ & $1$ & $0.5$ & $3\pi/2$ & $0.30$ & $18$ & $0.5$ & $0.5$ & $\pi/2$ & $0$ \\
$3$ & $1$ & $0.5$ & $\pi$ & $0$ & $11$ & $1$ & $0.5$ & $3\pi/2$ & $0.45$ & $19$ & $1.5$ & $0.5$ & $\pi/2$ & $0$ \\
$4$ & $1$ & $0.5$ & $3\pi/2$ & $0$ & $12$ & $1$ & $0.5$ & $3\pi/2$ & $0.60$ & \\
$5$ & $1$ & $0.5$ & $\pi/2$ & $0.15$ & $13$ & $1$ & $0.25$ & $\pi/2$ & $0$ & \\
$6$ & $1$ & $0.5$ & $\pi/2$ & $0.30$ & $14$ & $1$ & $0.75$ & $\pi/2$ & $0$ & \\
$7$ & $1$ & $0.5$ & $\pi/2$ & $0.45$ & $15$ & $1$ & $1$ & $\pi/2$ & $0$ & \\
$8$ & $1$ & $0.5$ & $\pi/2$ & $0.60$ & $16$ & $1$ & $1.25$ & $\pi/2$ & $0$ & \\

\bottomrule
\end{tabular}
\end{table}


\section{Results}\label{sec:parametric}

\subsection{Vertical and spanwise vortex gust encounters}\label{subsec:phi}

\begin{figure}[htbp]
    \centering
    \begin{overpic}[width=1.0\linewidth]{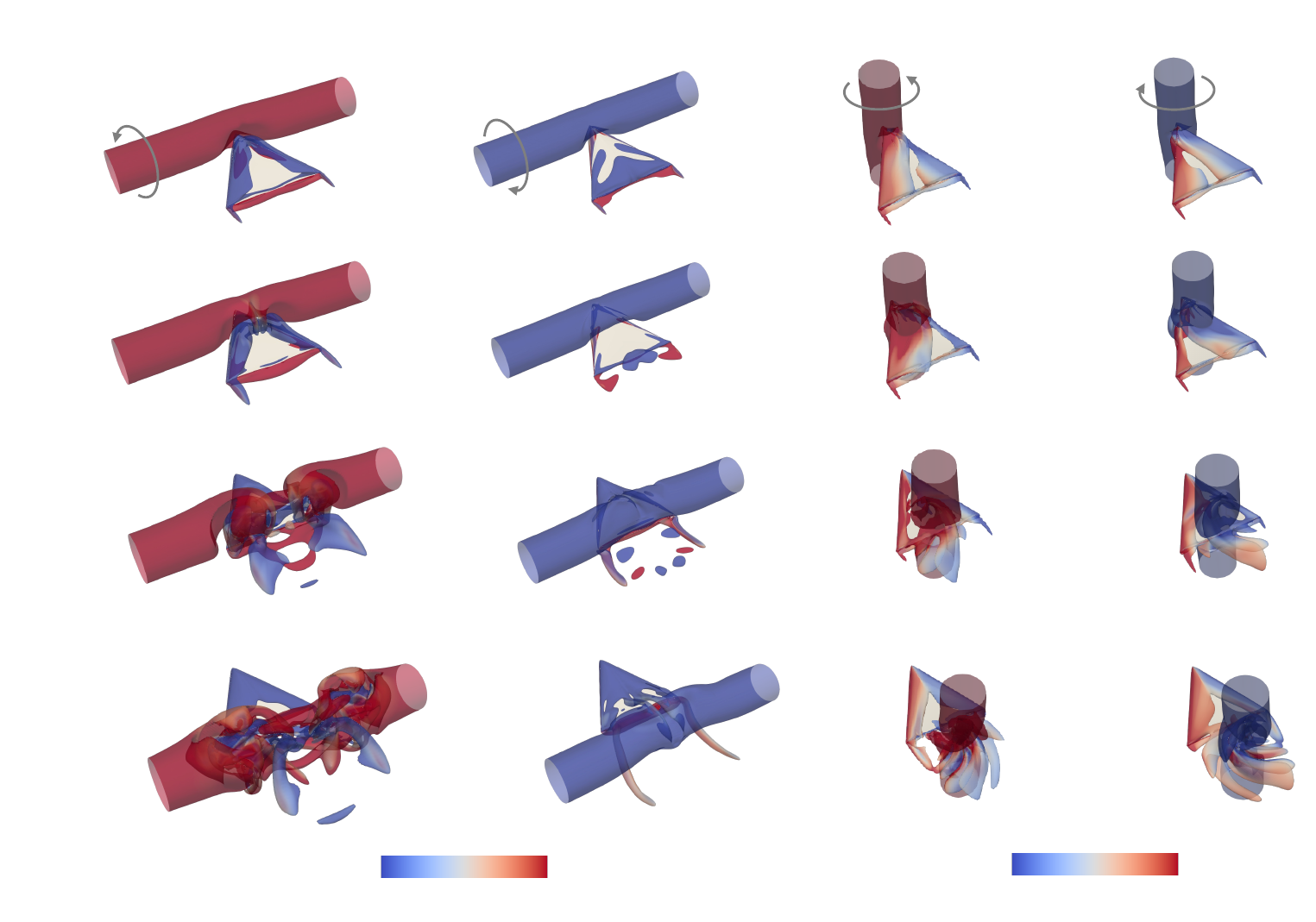}
    \put(0,56){{$t=0.0$}}
    \put(0,42.5){{$t=0.4$}}
    \put(0,29.5){{$t=0.9$}}
    \put(0,15){{$t=1.4$}}
    \put(15,67){{$\phi=0$}}
    \put(42,67){{$\phi=\pi$}}
    \put(65,67){{$\phi=\pi/2$}}
    \put(87.5,67){{$\phi=3\pi/2$}}
    \put(82.5,1.5){{$\omega_y$}}
    \put(75,5.7){{$-3$}}
    \put(88.9,5.7){{$3$}}
    \put(34.5,1.5){{$\omega_z$}}
    \put(27,5.7){{$-3$}}
    \put(41,5.7){{$3$}}
    \linethickness{1pt}
    \put(24,40){\color{black}\vector(-0.5,0.5){5}}
    \put(25,39){{arch vortex}}
    \put(26.1,10){\color{black}\vector(-0.7,-0.1){3.82}}
    \put(26.9,9.4){{hairpin vortex}}
    \put(52.4,10.5){\color{black}\vector(0.4,0.5){2}}
    \put(51,8.9){{hairpin vortex}}
    \put(78,15){{lobed vortex}}
    \put(79,14){\color{black}\vector(-0.4,-0.3){4}}
    \put(87,14){\color{black}\vector(0.4,-0.15){6}}
    \put(77,32){{lobed vortex}}
    \put(78,31){\color{black}\vector(-0.4,-0.3){5}}
    \put(86,31){\color{black}\vector(0.4,-0.14){8.9}}
    \end{overpic}
    \caption{the instantaneous $Q$ isosurface coloured by spanwise or wall-normal vorticity for cases with different gust orientation $\phi$ corresponding to case IDs $1$ through $4$ ($G=1$ and $D=0.5$).}
    \label{fig:phi_Q}
\end{figure}
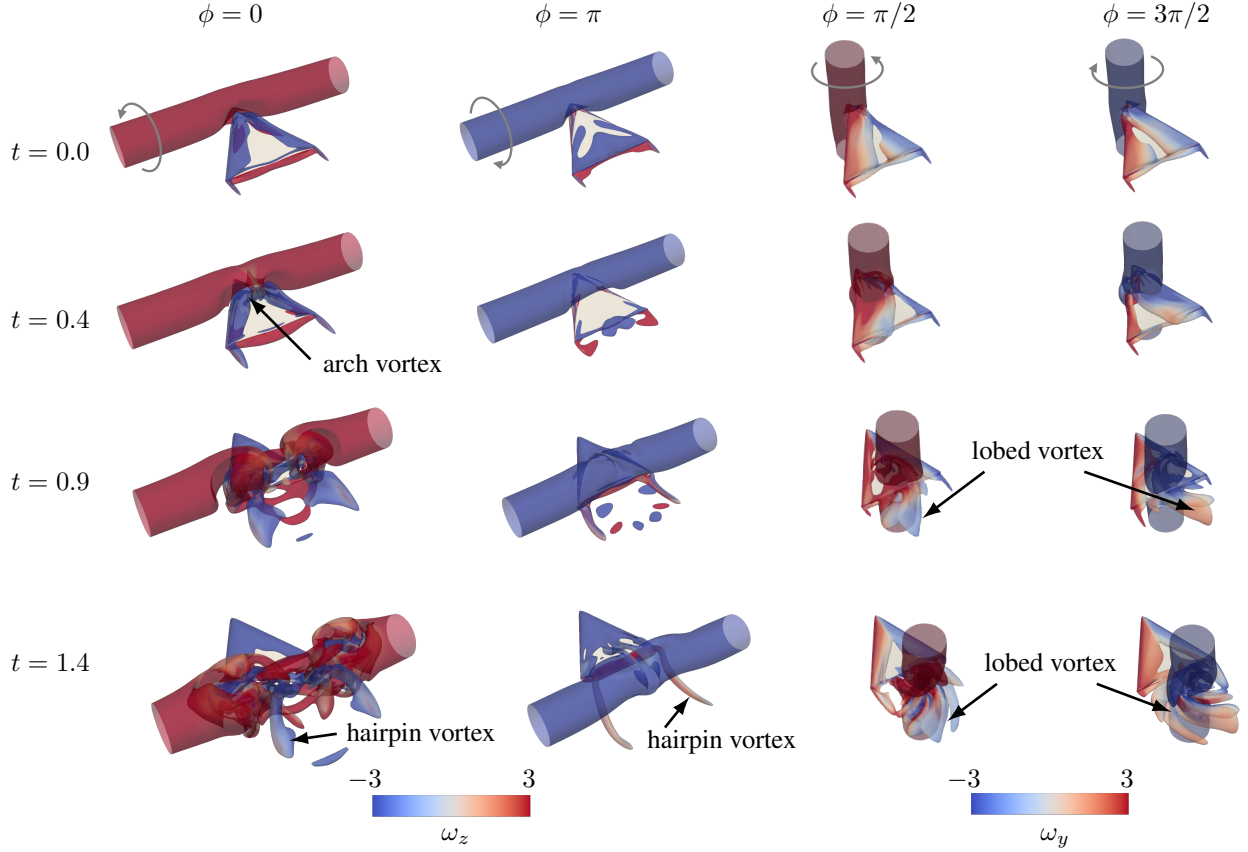

Let us begin by examining the difference in the flow response between spanwise and vertical vortex gust encounters. The cases examined here have case IDs $1$ through $4$, as summarized in table \ref{tab:gust_parameters}. The instantaneous $Q$-criterion isosurface is shown in figure \ref{fig:phi_Q}. We observe that the flow structures induced by the spanwise gusts (i.e., $\phi=0$ and $\pi$) are both symmetric about the root plane. For $\phi=0$, we observe that the preexisting boundary layer is lifted from the top surface of the delta wing at $t=0$, accompanied by a stronger leading-edge vortex, due to the upwash induced by the gust. The lifted boundary layer interacts with the gust and convects downstream, becoming an arch vortex at $t=0.4$. When the gust convects close to trailing edge, at $t=0.9$, the tip vortex is enhanced and lifted due to the upwash induced by the gust. In the meantime, the trailing-edge vortex is induced and shed downstream. After the gust passes the delta wing, at $t = 1.4$, we observe a hairpin vortex with head originating from the leading-edge vortex and legs originating from the tip vortex. 

The $\phi=\pi$ gust also induces a symmetric response, but from a downwash on the wing at $t=0$. As the gust convects downstream, the tip vortex is enhanced and pulled downward, as $t=0.9$ indicates. A trailing-edge vortex is formed due to the upwash after the gust passes the delta wing at $t=1.4$, and we can observe a primary hairpin vortex with heads originating from the trailing-edge vortex and legs originating from the tip vortex. This later upwash also lifts the recovered boundary layer, so there is a secondary hairpin vortex observed as well. A similar flow structure was observed by Odaka et al. \cite{odaka_square_2026,Odaka2026}, where they studied the spanwise vortex gust encounters of a rectangular wing, and readers are referred to these prior works for more detail.

The flow response of the two vertical vortex gust encounters (i.e., $\phi=\pi/2$ and $3\pi/2$) contrast sharply with those of the spanwise encounters. The $Q$ isosurfaces of these vertical cases are mirrors of each other, as expected. For either case, in the later stages of the encounter at $t = 0.9$ and $1.4$, we observe lobed vortical structures that twist around the gust vortex tube. These structures were not observed during the spanwise vortex gust encounters, but in section \ref{subsec:force_element} we will show that they make a critical contribution to the aerodynamic lift.

\begin{figure}[htbp]
    \centering
    \begin{overpic}[width=0.75\linewidth]{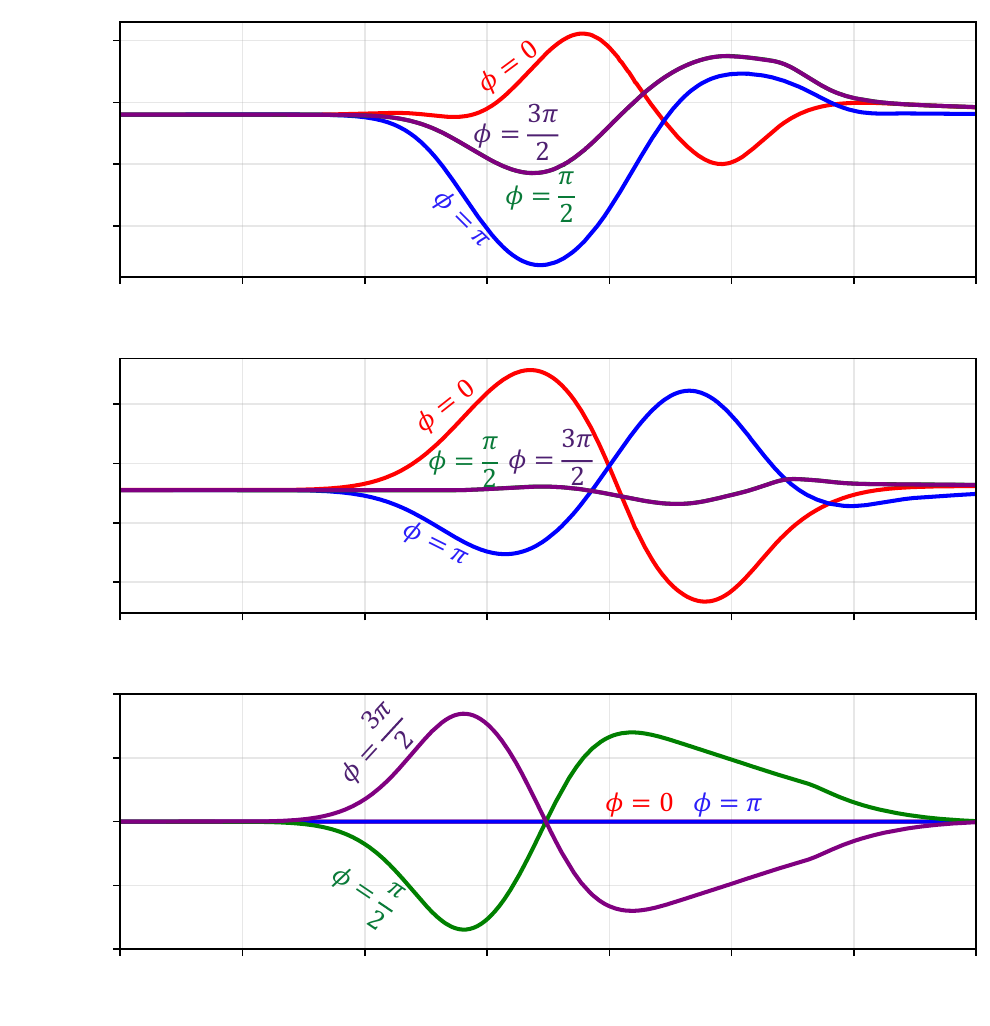}
    \put(8.6,4.2){\small{$-1.5$}}
    \put(20.5,4.2){\small{$-1.0$}}
    \put(32.3,4.2){\small{$-0.5$}}
    \put(46,4.2){\small{$0.0$}}
    \put(57.7,4.2){\small{$0.5$}}
    \put(69.8,4.2){\small{$1.0$}}
    \put(81.5,4.2){\small{$1.5$}}
    \put(93.4,4.2){\small{$2.0$}}
    
    \put(8.6,36.8){\small{$-1.5$}}
    \put(20.5,36.8){\small{$-1.0$}}
    \put(32.3,36.8){\small{$-0.5$}}
    \put(46,36.8){\small{$0.0$}}
    \put(57.7,36.8){\small{$0.5$}}
    \put(69.8,36.8){\small{$1.0$}}
    \put(81.5,36.8){\small{$1.5$}}
    \put(93.4,36.8){\small{$2.0$}}

    \put(8.6,69.7){\small{$-1.5$}}
    \put(20.5,69.7){\small{$-1.0$}}
    \put(32.3,69.7){\small{$-0.5$}}
    \put(46,69.7){\small{$0.0$}}
    \put(57.7,69.7){\small{$0.5$}}
    \put(69.8,69.7){\small{$1.0$}}
    \put(81.5,69.7){\small{$1.5$}}
    \put(93.4,69.7){\small{$2.0$}}

    \put(6.8,95.3){\small{$0.3$}}
    \put(6.8,89.2){\small{$0.2$}}
    \put(6.8,83.2){\small{$0.1$}}
    \put(6.8,77.2){\small{$0.0$}}

    \put(6.8,59.9){\small{$1.0$}}
    \put(6.8,54){\small{$0.5$}}
    \put(6.8,48.3){\small{$0.0$}}
    \put(4.8,42.5){\small{$-0.5$}}

    \put(5.8,31.5){\small{$0.10$}}
    \put(5.8,25.4){\small{$0.05$}}
    \put(5.8,19.2){\small{$0.00$}}
    \put(3.8,12.9){\small{$-0.05$}}
    \put(3.8,6.8){\small{$-0.10$}}

    \put(0,82.5){\rotatebox{90}{\large$C_D$}}
    \put(0,50.5){\rotatebox{90}{\large$C_L$}}
    \put(0,19){\rotatebox{90}{\large$C_S$}}

    \put(52,0.5){{\large$t$}}
    \end{overpic}
    \caption{the aerodynamic force histories during gust encounters with different gust orientation $\phi$.}
    \label{fig:phi_aerodynamics}
\end{figure}
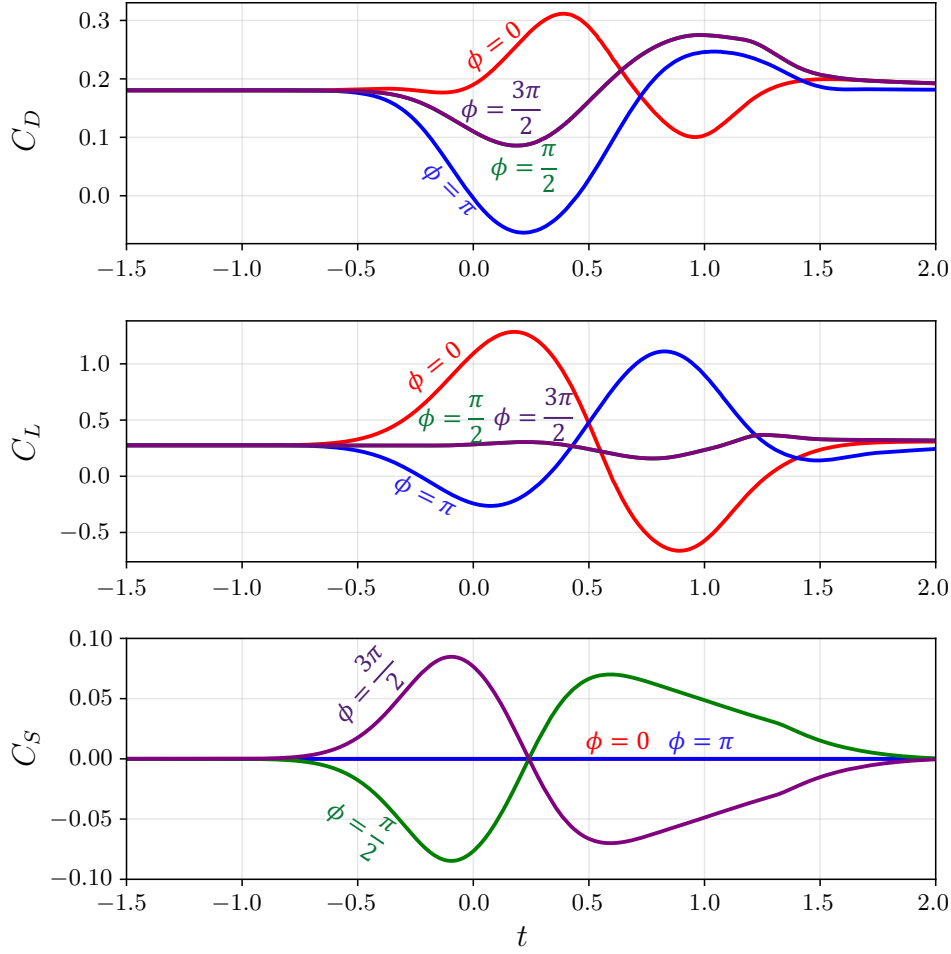

Figure \ref{fig:phi_aerodynamics} depicts the three components of aerodynamic force coefficient for these different gust orientations. For the two spanwise vortex gust encounters, we can see that only the drag coefficient $C_D$ and lift coefficient $C_L$ are disrupted, while the side force coefficient $C_S$ remains zero due to the symmetric configuration. However, the two vertical vortex gust encounters show that all three aerodynamic force components are disrupted, though with lift variations that are less substantial than the spanwise gust cases. As expected from the mirror symmetry about the root plane of these two vertical cases, their drag and lift histories are identical and their side force histories are equal and opposite. Furthermore, these vertical vortex cases show that the side force responds earlier than the other components.

Much of the effort of the remainder of this paper will be devoted to revealing the physical processes that explain the formation and evolution of the lobed vortical structures that emerge from vertical vortex encounters and identify their contribution to the aerodynamic loads.


\subsection{Dynamics of vertical vortex gust encounters}\label{subsec:z0}

\subsubsection{Flow structures and the aerodynamic response}\label{subsec:bigpicture}

The goal of this section is to illuminate the physical processes that govern vertical vortex encounters with a tailless delta wing. First, we will examine the overall behavior of this encounter---the induced large-scale flow structures and global aerodynamic response---over a range of different lateral (i.e., spanwise) positions of the gust. We will then analyze representative cases in more detail in the following sections.

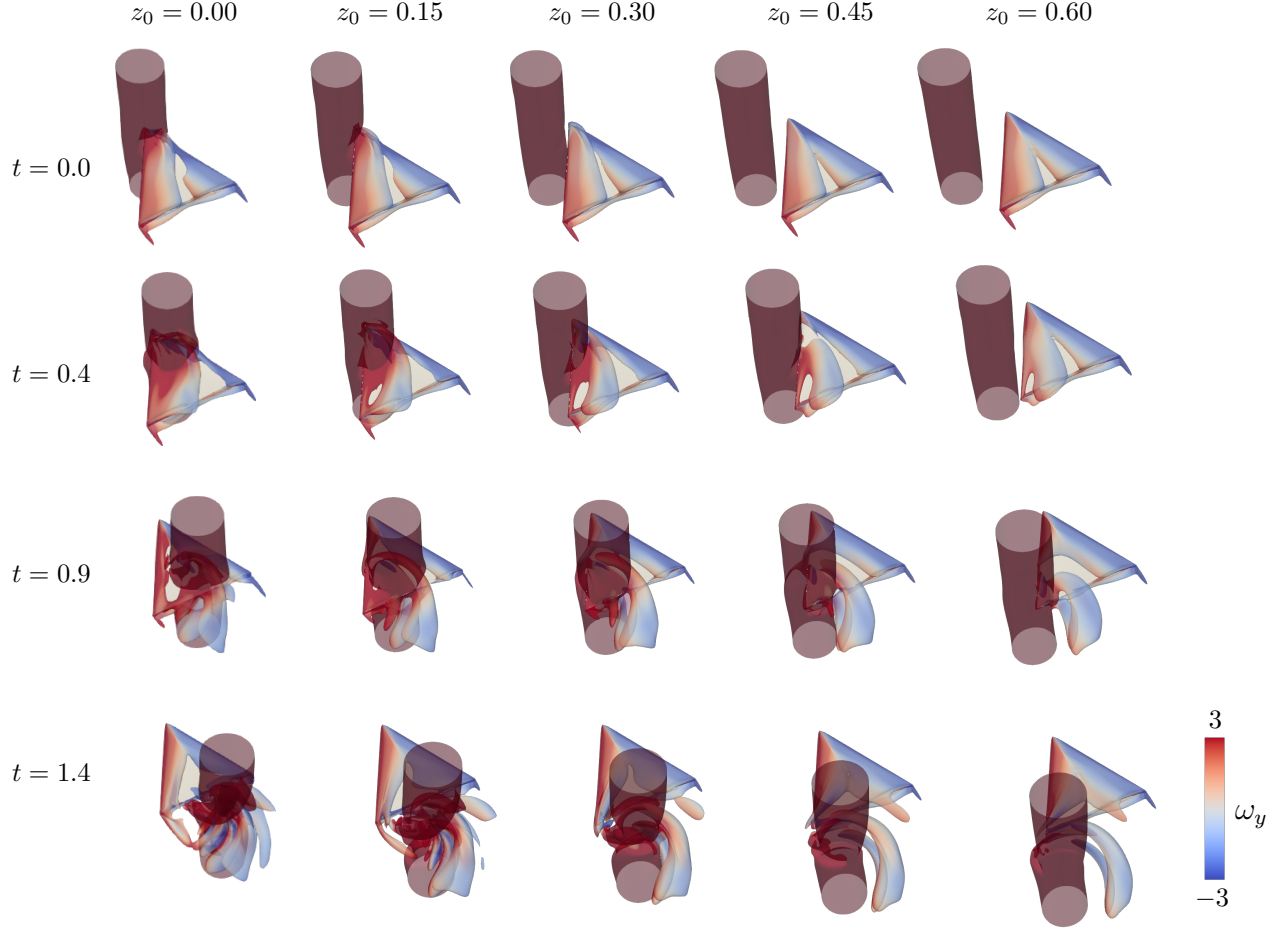
\begin{figure}[htbp]
    \centering
    \begin{overpic}[width=1.0\linewidth]{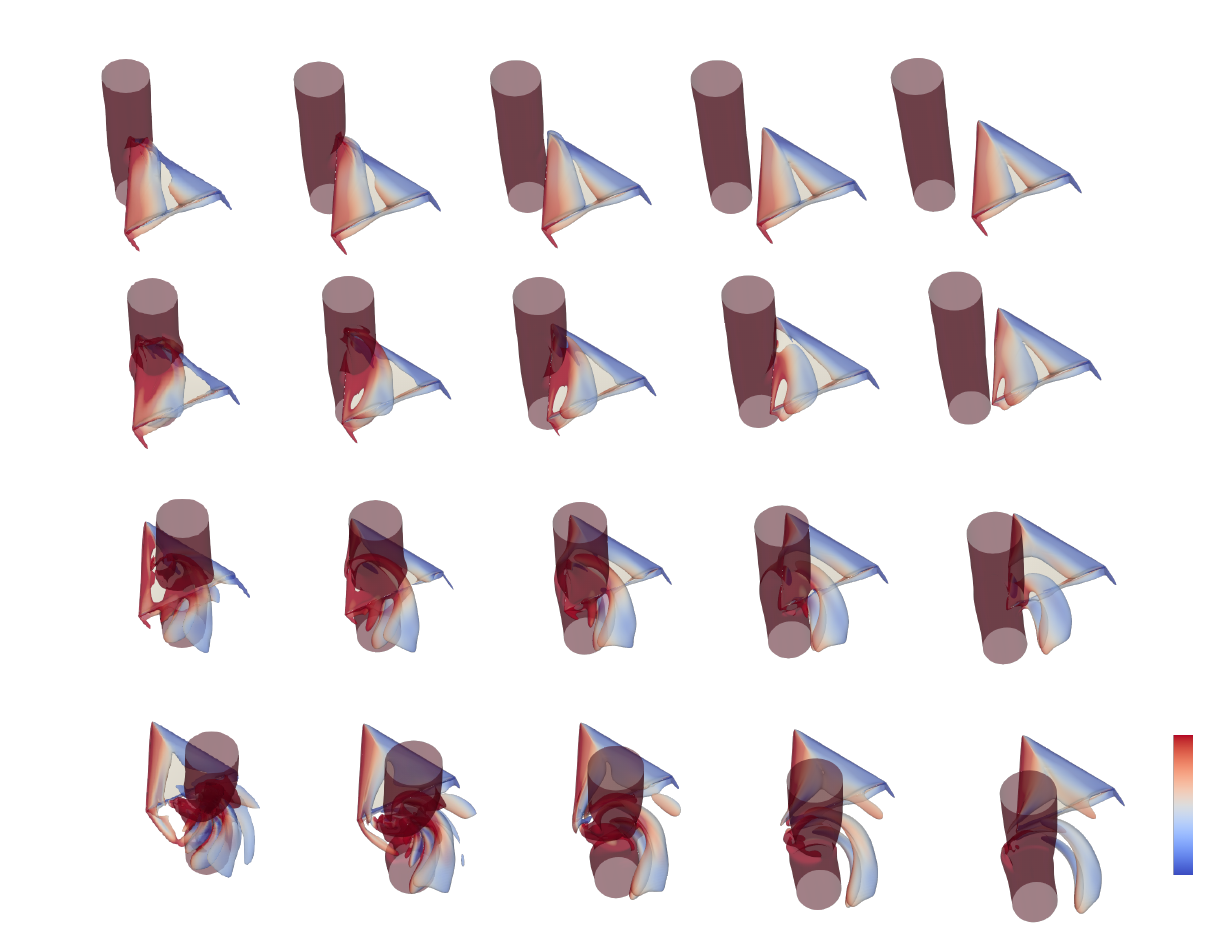}
    \put(0,61.5){{$t=0.0$}}
    \put(0,45){{$t=0.4$}}
    \put(0,29){{$t=0.9$}}
    \put(0,13){{$t=1.4$}}
    \put(9.5,74){{$z_0=0.00$}}
    \put(26,74){{$z_0=0.15$}}
    \put(43,74){{$z_0=0.30$}}
    \put(60.5,74){{$z_0=0.45$}}
    \put(78,74){{$z_0=0.60$}}
    \put(98,10){{\large$\omega_y$}}
    \put(94.8,3){{$-3$}}
    \put(95.9,17.3){{$3$}}
    \end{overpic}
    \caption{the instantaneous $Q$ isosurface coloured by wall-normal vorticity $\omega_y$ for counter-clockwise vertical vortex gust encounters ($\phi=\pi/2$) with different lateral position $z_0$.}
    \label{fig:Q_z0_p}
\end{figure}

The evolution of the $Q$ isosurface is shown in figure \ref{fig:Q_z0_p} for encounters of a counter-clockwise vertical vortex gust (i.e. $\phi=\pi/2$) whose lateral position is varied from the root to the left tip. The cases considered here correspond to case IDs $2,5,6,7,8$ in table \ref{tab:gust_parameters}. The disturbance of the wing's flow is delayed as the gust shifts laterally from root to tip as we can see from the results at $t=0.0$, at which time the boundary layer is distorted in the root-centered ($z_0=0$) gust encounter, but is increasingly less distorted as the gust shifts laterally until it remains nearly intact for the most outboard encounter at $z_0=0.6$. Furthermore, the emergent lobed vortical structures that twist around the gust vortex tube can be observed by $t = 1.4$ for all five cases, but in reduced number in the most outboard gust compared to the root-centered gust encounter. The gust's influence on the wing becomes increasingly marginalized as the vortex position shifts laterally, so that the opposite side of the wing remains nearly unaffected for the most outboard vortex encounter.

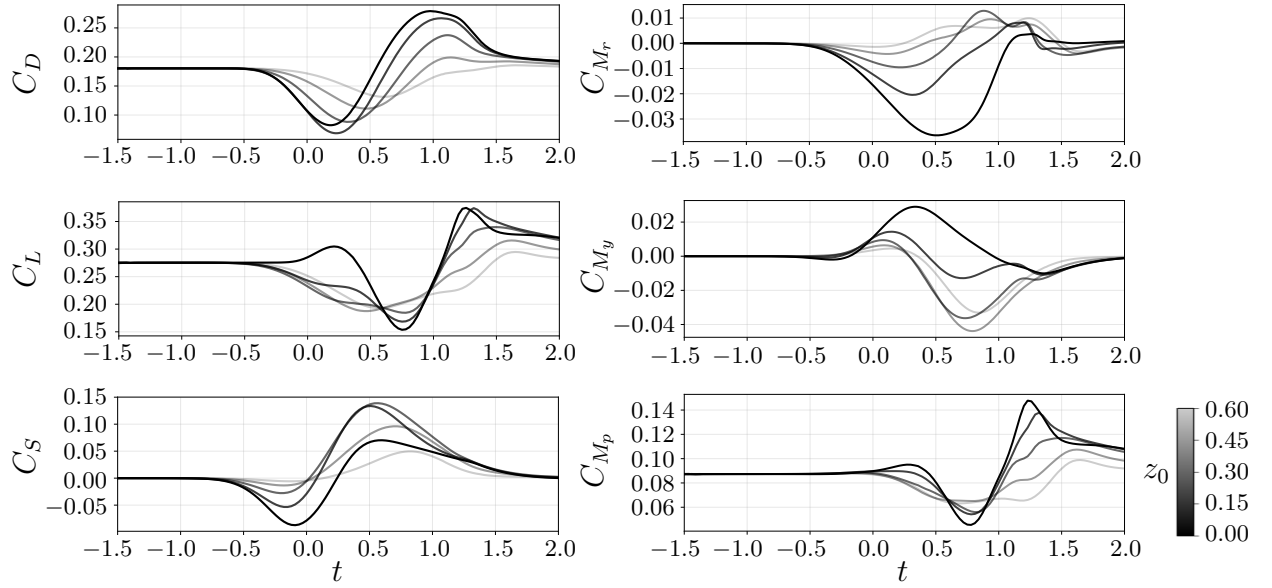
\begin{figure}[htbp]
    \centering
    \begin{overpic}[width=1.0\linewidth]{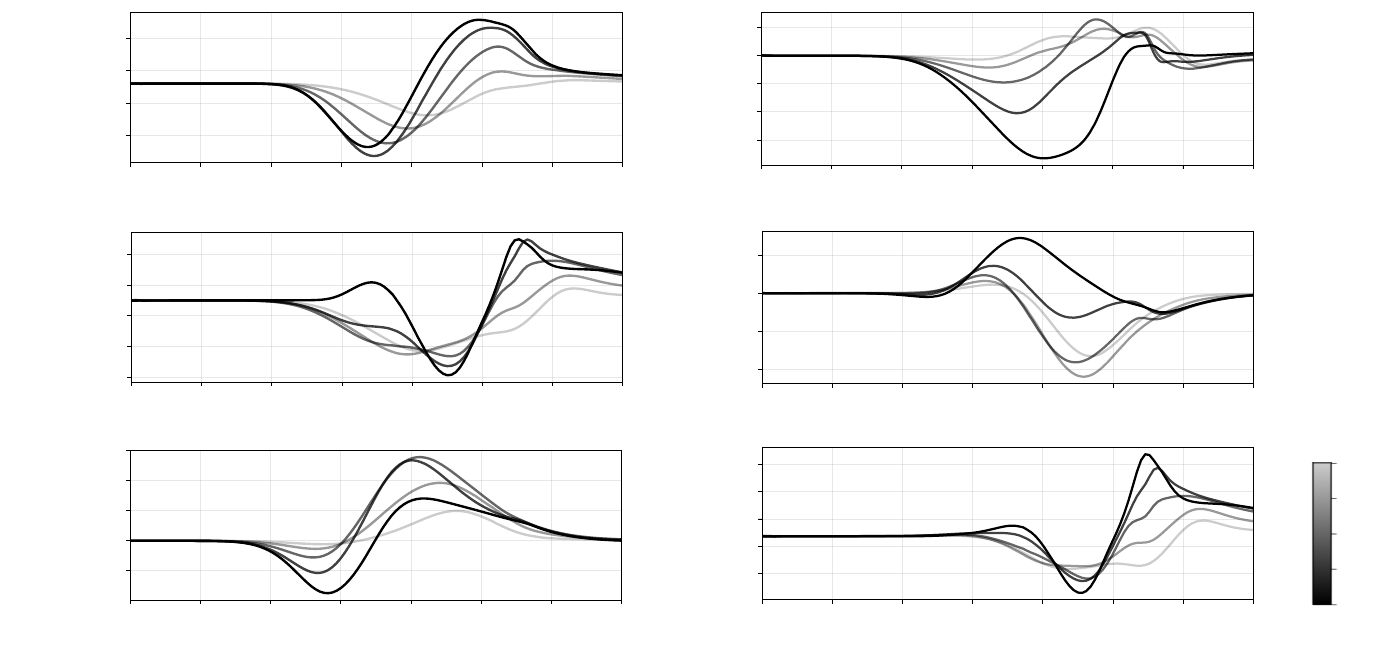}
    \put(6.5,2){\small{$-1.5$}}
    \put(11.6,2){\small{$-1.0$}}
    \put(16.8,2){\small{$-0.5$}}
    \put(23.4,2){\small{$0.0$}}
    \put(28.6,2){\small{$0.5$}}
    \put(33.5,2){\small{$1.0$}}
    \put(38.6,2){\small{$1.5$}}
    \put(43.6,2){\small{$2.0$}}

    \put(6.5,17.7){\small{$-1.5$}}
    \put(11.6,17.7){\small{$-1.0$}}
    \put(16.8,17.7){\small{$-0.5$}}
    \put(23.4,17.7){\small{$0.0$}}
    \put(28.6,17.7){\small{$0.5$}}
    \put(33.5,17.7){\small{$1.0$}}
    \put(38.6,17.7){\small{$1.5$}}
    \put(43.6,17.7){\small{$2.0$}}

    \put(6.5,33.5){\small{$-1.5$}}
    \put(11.6,33.5){\small{$-1.0$}}
    \put(16.8,33.5){\small{$-0.5$}}
    \put(23.4,33.5){\small{$0.0$}}
    \put(28.6,33.5){\small{$0.5$}}
    \put(33.5,33.5){\small{$1.0$}}
    \put(38.6,33.5){\small{$1.5$}}
    \put(43.6,33.5){\small{$2.0$}}

    \put(52,2){\small{$-1.5$}}
    \put(57,2){\small{$-1.0$}}
    \put(62.1,2){\small{$-0.5$}}
    \put(68.7,2){\small{$0.0$}}
    \put(73.8,2){\small{$0.5$}}
    \put(78.8,2){\small{$1.0$}}
    \put(83.9,2){\small{$1.5$}}
    \put(88.9,2){\small{$2.0$}}

    \put(52,17.5){\small{$-1.5$}}
    \put(57,17.5){\small{$-1.0$}}
    \put(62.1,17.5){\small{$-0.5$}}
    \put(68.7,17.5){\small{$0.0$}}
    \put(73.8,17.5){\small{$0.5$}}
    \put(78.8,17.5){\small{$1.0$}}
    \put(83.9,17.5){\small{$1.5$}}
    \put(88.9,17.5){\small{$2.0$}}

    \put(52,33.2){\small{$-1.5$}}
    \put(57,33.2){\small{$-1.0$}}
    \put(62.1,33.2){\small{$-0.5$}}
    \put(68.7,33.2){\small{$0.0$}}
    \put(73.8,33.2){\small{$0.5$}}
    \put(78.8,33.2){\small{$1.0$}}
    \put(83.9,33.2){\small{$1.5$}}
    \put(88.9,33.2){\small{$2.0$}}

    \put(3.5,5.4){\small{$-0.05$}}
    \put(5,7.5){\small{$0.00$}}
    \put(5,9.8){\small{$0.05$}}
    \put(5,12){\small{$0.10$}}
    \put(5,14.2){\small{$0.15$}}

    \put(5,19.5){\small{$0.15$}}
    \put(5,21.7){\small{$0.20$}}
    \put(5,23.9){\small{$0.25$}}
    \put(5,26.1){\small{$0.30$}}
    \put(5,28.3){\small{$0.35$}}

    \put(5,36.8){\small{$0.10$}}
    \put(5,39.2){\small{$0.15$}}
    \put(5,41.6){\small{$0.20$}}
    \put(5,44){\small{$0.25$}}

    \put(50.5,5.3){\small{$0.06$}}
    \put(50.5,7.25){\small{$0.08$}}
    \put(50.5,9.2){\small{$0.10$}}
    \put(50.5,11.15){\small{$0.12$}}
    \put(50.5,13.1){\small{$0.14$}}

    \put(49,20){\small{$-0.04$}}
    \put(49,22.8){\small{$-0.02$}}
    \put(50.5,25.5){\small{$0.00$}}
    \put(50.5,28.2){\small{$0.02$}}

    \put(49,36.4){\small{$-0.03$}}
    \put(49,38.45){\small{$-0.02$}}
    \put(49,40.5){\small{$-0.01$}}
    \put(50.5,42.55){\small{$0.00$}}
    \put(50.5,44.6){\small{$0.01$}}

    \put(1,39){\rotatebox{90}{\large$C_D$}}
    \put(1,23.5){\rotatebox{90}{\large$C_L$}}
    \put(1,8.5){\rotatebox{90}{\large$C_S$}}

    \put(46.5,39){\rotatebox{90}{\large$C_{M_r}$}}
    \put(46.5,23){\rotatebox{90}{\large$C_{M_y}$}}
    \put(46.5,7.5){\rotatebox{90}{\large$C_{M_p}$}}

    \put(26.5,0){{\large$t$}}
    \put(71.8,0){{\large$t$}}

    \put(91.5,8.3){{\large$z_0$}}
    \put(96.5,3.2){{\small$0.00$}}
    \put(96.5,5.7){{\small$0.15$}}
    \put(96.5,8.2){{\small$0.30$}}
    \put(96.5,10.7){{\small$0.45$}}
    \put(96.5,13.2){{\small$0.60$}}
    \end{overpic}
    \caption{the aerodynamic force and moment histories during counter-clockwise vertical vortex gust encounters ($\phi=\pi/2$) with different lateral position $z_0$.}
    \label{fig:aero_z0_p}
\end{figure}

The associated aerodynamic response profiles for the five cases are shown in figure \ref{fig:aero_z0_p}. We first observe that all six components of the aerodynamic response are disrupted, with the side force and the roll and yaw moments responding earlier than the other three components. The drag and side force generally decrease as the gust approaches the wing, then increase as the gust approaches the trailing edge ($t = 1$), though the initial decrease in both is reduced for the most outboard gusts. Disturbances to the pitching moment are somewhat stronger than rolling and yawing moments. Furthermore, as the gust encounter shifts toward the tip, the aerodynamic response in all components becomes less severe, consistent with the previous observation that the gust-wing interaction becomes more marginalized.


\begin{figure}[htbp]
    \centering
    \begin{overpic}[width=1.0\linewidth]{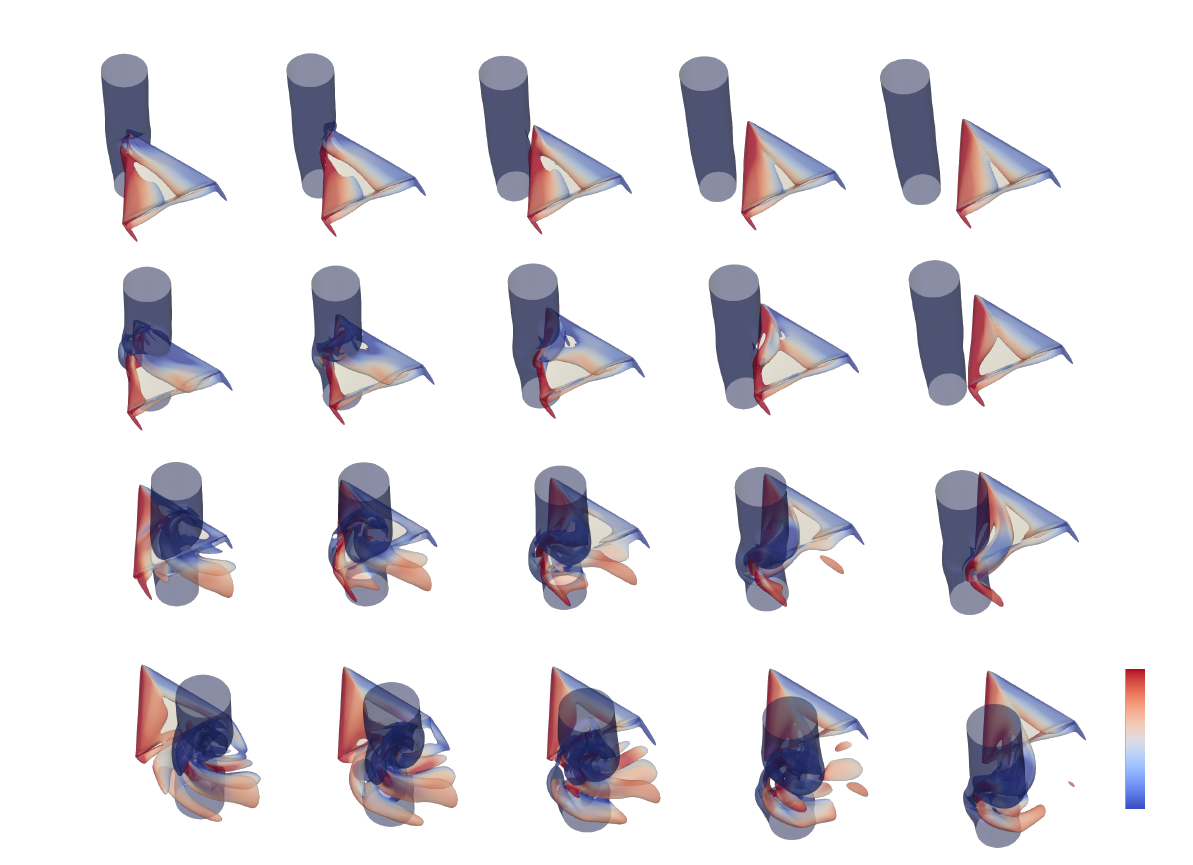}
    \put(0,57){{$t=0.0$}}
    \put(0,41.5){{$t=0.4$}}
    \put(0,27){{$t=0.9$}}
    \put(0,12){{$t=1.4$}}
    \put(9,70.2){{$z_0=0.00$}}
    \put(26,70.2){{$z_0=0.15$}}
    \put(43,70.2){{$z_0=0.30$}}
    \put(61.5,70.2){{$z_0=0.45$}}
    \put(79.5,70.2){{$z_0=0.60$}}
    \put(97,9.8){\large{$\omega_y$}}
    \put(93.7,2.6){{$-3$}}
    \put(94.7,17){{$3$}}
    \end{overpic}
    \caption{the instantaneous $Q$ isosurface coloured by wall-normal vorticity $\omega_y$ for clockwise vertical vortex gust encounters ($\phi=3\pi/2$) with different lateral position $z_0$.}
    \label{fig:Q_z0_n}
\end{figure}

We now examine the effect of the lateral position of a clockwise vertical vortex gust (i.e., $\phi=3\pi/2$), which corresponds to the case IDs $4,9,10,11,12$ in table \ref{tab:gust_parameters}. We note that, throughout the study, we use the word `left' to refer to the port side, and the word `right' to refer to the starboard side. We also note that varying a clockwise vortex's position from root to left tip, as we do here, is equivalent (modulo a reflection about the root) to varying a counter-clockwise vortex's position from the root to right tip: they produce the same drag and lift force and pitching moment and opposite side force, roll and yaw moment. In other words, the cases studied in this section comprise a tip to tip sweep of lateral position. The instantaneous $Q$ isosurfaces for this clockwise vortex's various lateral positions are shown in figure \ref{fig:Q_z0_n}. Similar to figure \ref{fig:Q_z0_p}, we can first observe at early times that the disturbance of the gust on the delta wing is delayed as the gust shifts laterally from root to tip. Also, we once again observe lobed vortical structures---though somewhat different in shape---twisting around the gust vortex tube for all five cases shown here, with fewer induced lobes and a more marginalized influence from the vortex as it shifts toward the outboard position.

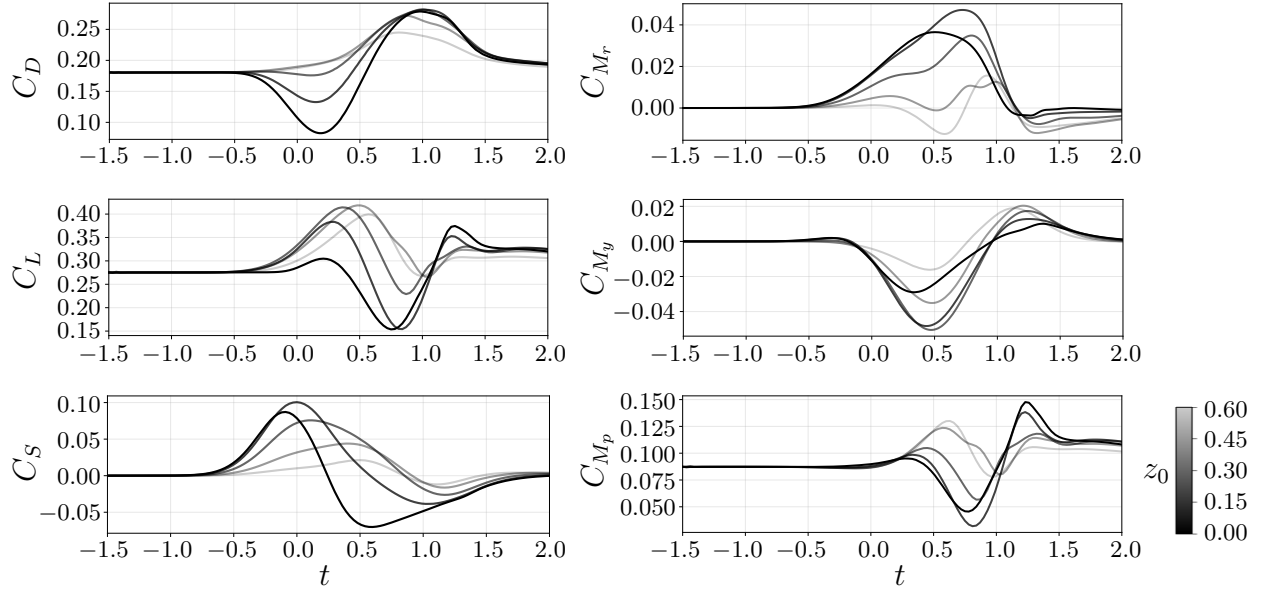
\begin{figure}[htbp]
    \centering
    \begin{overpic}[width=1.0\linewidth]{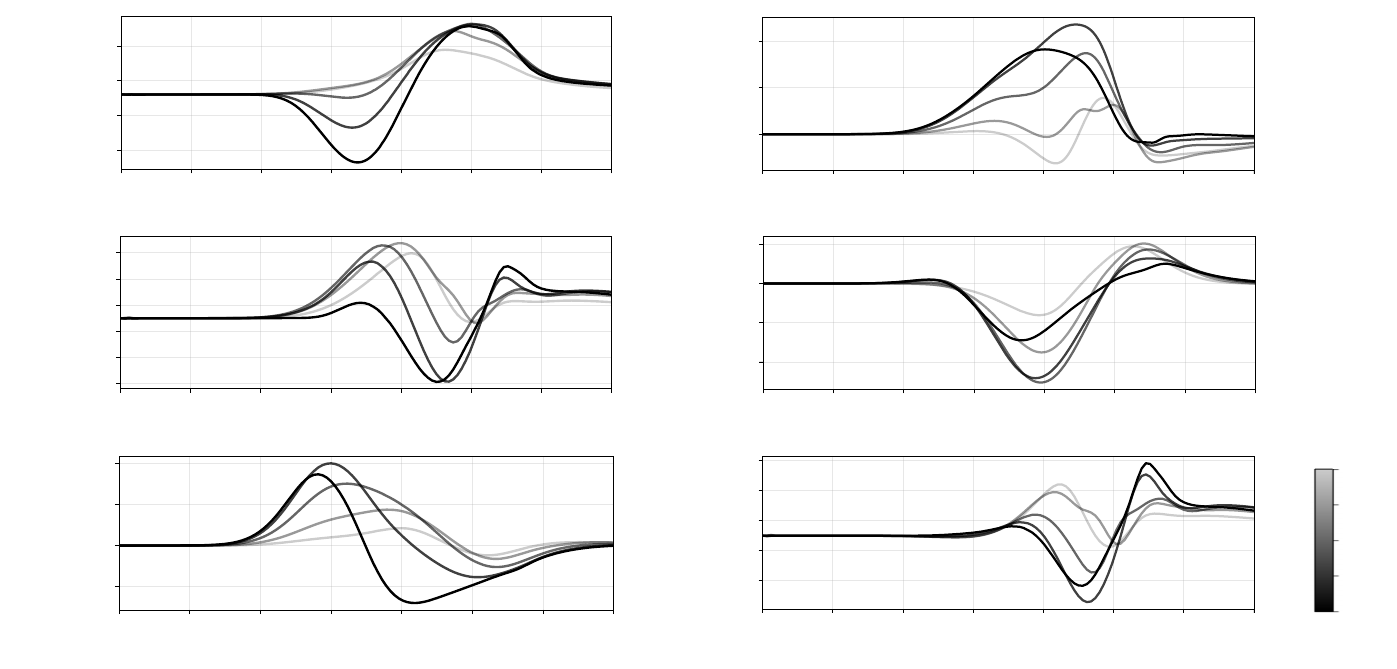}
    \put(6.2,2.3){\small{$-1.5$}}
    \put(11.3,2.3){\small{$-1.0$}}
    \put(16.5,2.3){\small{$-0.5$}}
    \put(22.6,2.3){\small{$0.0$}}
    \put(27.7,2.3){\small{$0.5$}}
    \put(32.8,2.3){\small{$1.0$}}
    \put(37.8,2.3){\small{$1.5$}}
    \put(42.8,2.3){\small{$2.0$}}

    \put(6.2,18){\small{$-1.5$}}
    \put(11.3,18){\small{$-1.0$}}
    \put(16.5,18){\small{$-0.5$}}
    \put(22.6,18){\small{$0.0$}}
    \put(27.7,18){\small{$0.5$}}
    \put(32.8,18){\small{$1.0$}}
    \put(37.8,18){\small{$1.5$}}
    \put(42.8,18){\small{$2.0$}}

    \put(6.2,33.8){\small{$-1.5$}}
    \put(11.3,33.8){\small{$-1.0$}}
    \put(16.5,33.8){\small{$-0.5$}}
    \put(22.6,33.8){\small{$0.0$}}
    \put(27.7,33.8){\small{$0.5$}}
    \put(32.8,33.8){\small{$1.0$}}
    \put(37.8,33.8){\small{$1.5$}}
    \put(42.8,33.8){\small{$2.0$}}

    \put(52,2.3){\small{$-1.5$}}
    \put(57,2.3){\small{$-1.0$}}
    \put(62.1,2.3){\small{$-0.5$}}
    \put(68.7,2.3){\small{$0.0$}}
    \put(73.8,2.3){\small{$0.5$}}
    \put(78.8,2.3){\small{$1.0$}}
    \put(83.9,2.3){\small{$1.5$}}
    \put(88.9,2.3){\small{$2.0$}}

    \put(52,18){\small{$-1.5$}}
    \put(57,18){\small{$-1.0$}}
    \put(62.1,18){\small{$-0.5$}}
    \put(68.7,18){\small{$0.0$}}
    \put(73.8,18){\small{$0.5$}}
    \put(78.8,18){\small{$1.0$}}
    \put(83.9,18){\small{$1.5$}}
    \put(88.9,18){\small{$2.0$}}

    \put(52,33.8){\small{$-1.5$}}
    \put(57,33.8){\small{$-1.0$}}
    \put(62.1,33.8){\small{$-0.5$}}
    \put(68.7,33.8){\small{$0.0$}}
    \put(73.8,33.8){\small{$0.5$}}
    \put(78.8,33.8){\small{$1.0$}}
    \put(83.9,33.8){\small{$1.5$}}
    \put(88.9,33.8){\small{$2.0$}}

    \put(3,5.2){\small{$-0.05$}}
    \put(4.5,8.1){\small{$0.00$}}
    \put(4.5,11){\small{$0.05$}}
    \put(4.5,13.9){\small{$0.10$}}

    \put(4.5,19.8){\small{$0.15$}}
    \put(4.5,21.7){\small{$0.20$}}
    \put(4.5,23.5){\small{$0.25$}}
    \put(4.5,25.4){\small{$0.30$}}
    \put(4.5,27.3){\small{$0.35$}}
    \put(4.5,29.2){\small{$0.40$}}

    \put(4.5,36.5){\small{$0.10$}}
    \put(4.5,39){\small{$0.15$}}
    \put(4.5,41.5){\small{$0.20$}}
    \put(4.5,44){\small{$0.25$}}

    \put(49.6,5.6){\small{$0.050$}}
    \put(49.6,7.7){\small{$0.075$}}
    \put(49.6,9.9){\small{$0.100$}}
    \put(49.6,12){\small{$0.125$}}
    \put(49.6,14.1){\small{$0.150$}}

    \put(49,21.3){\small{$-0.04$}}
    \put(49,24){\small{$-0.02$}}
    \put(50.5,26.8){\small{$0.00$}}
    \put(50.5,29.6){\small{$0.02$}}

    \put(50.5,37.7){\small{$0.00$}}
    \put(50.5,41){\small{$0.02$}}
    \put(50.5,44.3){\small{$0.04$}}

    \put(1,39){\rotatebox{90}{\large$C_D$}}
    \put(1,23.7){\rotatebox{90}{\large$C_L$}}
    \put(1,8.3){\rotatebox{90}{\large$C_S$}}

    \put(46.5,39){\rotatebox{90}{\large$C_{M_r}$}}
    \put(46.5,23.2){\rotatebox{90}{\large$C_{M_y}$}}
    \put(46.5,7.8){\rotatebox{90}{\large$C_{M_p}$}}

    \put(25.5,0){{\large$t$}}
    \put(71.6,0){{\large$t$}}

    \put(91.5,8.6){{\large$z_0$}}
    \put(96.4,3.6){{\small$0.00$}}
    \put(96.4,6.1){{\small$0.15$}}
    \put(96.4,8.6){{\small$0.30$}}
    \put(96.4,11.1){{\small$0.45$}}
    \put(96.4,13.5){{\small$0.60$}}
    \end{overpic}
    \caption{the aerodynamic force and moment histories during clockwise vertical vortex gust encounters ($\phi=3\pi/2$) with different lateral position $z_0$.}
    \label{fig:aero_z0_n}
\end{figure}

The associated force and moment response for these clockwise gusts is shown in figure \ref{fig:aero_z0_n}. We can make a few observations similar to those we made in figure \ref{fig:aero_z0_p}. First, all six components of the aerodynamic response are disrupted with the side force and the roll and yaw moments responding earlier than the other components. The disturbance to pitching moment is once again strongest of all moment components. And again, as the gust shifts laterally from root to tip, the aerodynamic response becomes less severe due to a more marginalized gust-wing interaction. However, there are some features that are unique to this set of cases. The drag force profiles for $z_0=0.45$ and $0.60$ show different trends compared to the corresponding counter-clockwise outboard cases: instead of exhibiting a decrease, the drag force now increases as the vortex approaches the wing. The trends in side force, however, are similar to those for the counter-clockwise cases, exhibiting the same mitigated influence on the initial deviation.  



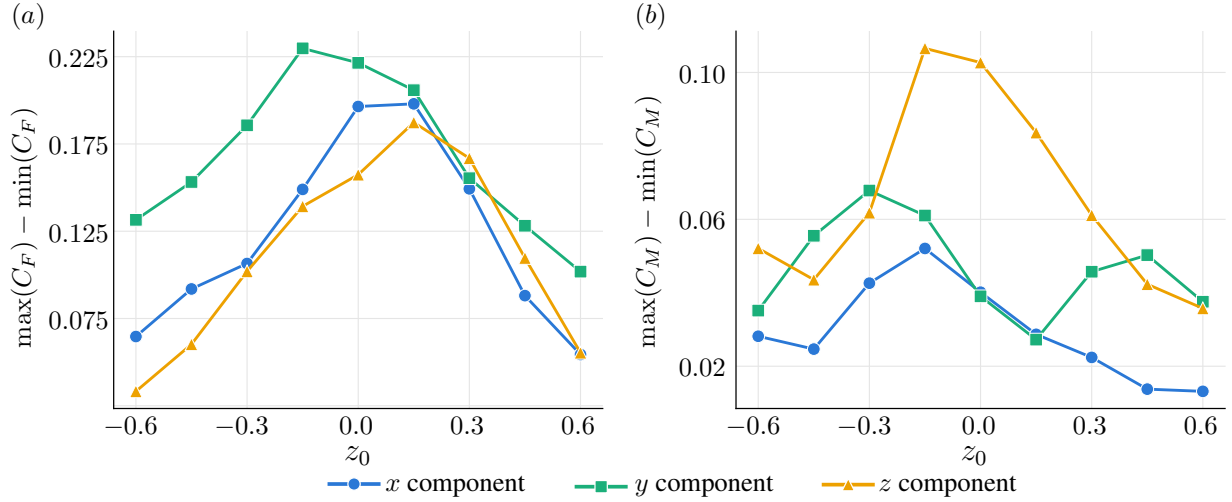
\begin{figure}[htbp]
    \centering
    \begin{overpic}[width=1.0\linewidth]{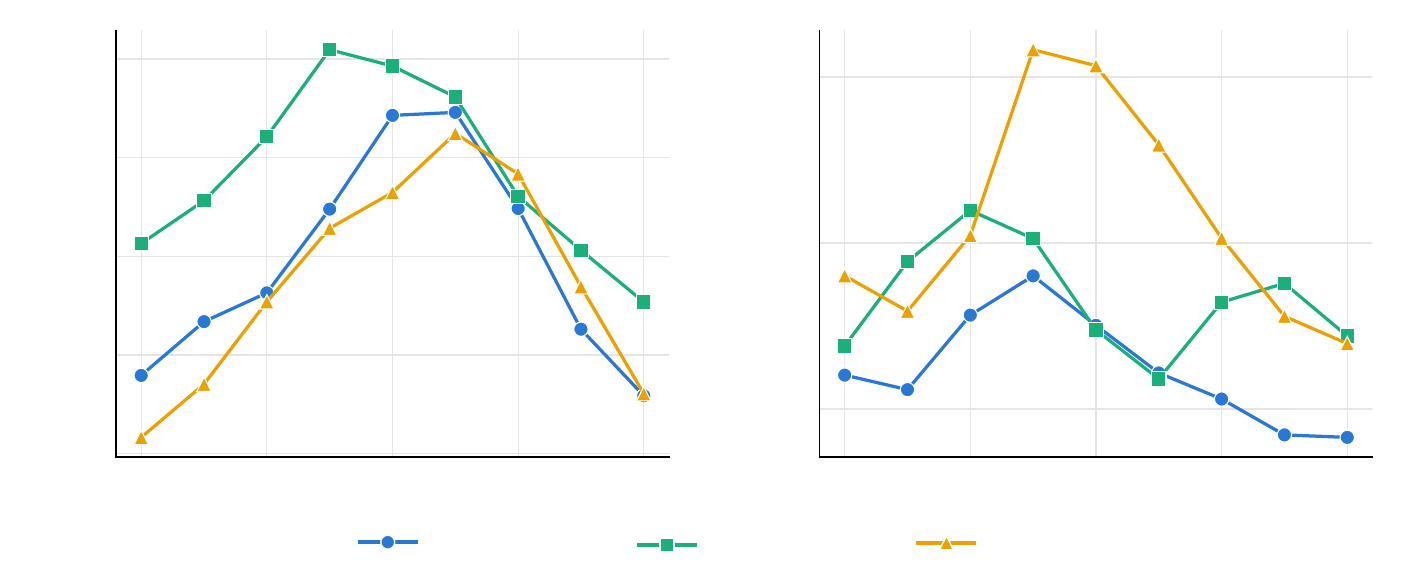}
    \put(0,39){{$(a)$}}
    \put(50,39){{$(b)$}}
    \put(26.7,4){\large{$z_0$}}
    \put(76.1,4){\large{$z_0$}}
    \put(7.4,6.1){{$-0.6$}}
    \put(16.3,6.1){{$-0.3$}}
    \put(26.3,6.1){{$0.0$}}
    \put(35.2,6.1){{$0.3$}}
    \put(44.1,6.1){{$0.6$}}
    \put(57.3,6.1){{$-0.6$}}
    \put(66.2,6.1){{$-0.3$}}
    \put(76.2,6.1){{$0.0$}}
    \put(85.1,6.1){{$0.3$}}
    \put(94,6.1){{$0.6$}}
    \put(3,14.5){{$0.075$}}
    \put(3,21.5){{$0.125$}}
    \put(3,28.5){{$0.175$}}
    \put(3,35.5){{$0.225$}}
    \put(0,13){\rotatebox{90}{$\max(C_F)-\min(C_F)$}}
    \put(53.5,10.7){{$0.02$}}
    \put(53.5,22.5){{$0.06$}}
    \put(53.5,34.2){{$0.10$}}
    \put(50,13){\rotatebox{90}{$\max(C_M)-\min(C_M)$}}
    \put(30,1.5){{$x$ component}}
    \put(50,1.5){{$y$ component}}
    \put(69.5,1.5){{$z$ component}}
    \end{overpic}
    \caption{The effect of the gust lateral position $z_0$ on the peak-to-peak magnitude of the disturbed $(a)$ aerodynamic force and $(b)$ aerodynamic moment.}
    \label{fig:peak_z}
\end{figure}

From this parametric sweep, we can address another question of practical relevance: how far the gust must pass from the wing, laterally, before its effect on the loads becomes insignificant. Therefore, we examine its effect on the disturbance magnitude of each aerodynamic response component in figure \ref{fig:peak_z}. Except for the yaw moment, all the other five components are strongest for near-root encounters (maximum at $|z_0|\leq0.15$) and decay gradually with offest, retaining roughly one-fifth to one-half maximal peak-to-peak response even when the gust center passes directly over the wingtip ($|z_0|=0.6$). The yaw moment, driven mostly by the asymmetry of the spanwise load distribution, does not follow the same pattern. It is most severe at the intermediate offsets where the gust impinges mainly on one half of the wing, and at the tip-centered encounter remains comparable to its value at the root-centered encounter.

With the examination of cases in this section, we have a big picture of how vertical vortex gust encounters induce flow structures and disrupt the aerodynamic response. We have observed some relatively simple trends in drag and side force, including some evidence of competition among the contributors to these components. However, the other components---lift force or moment---exhibit more complex dependence on lateral position, indicating that there are multiple influences on these components. We have also seen the emergence of lobed vortex structures that wrap around the gust vortex. Therefore, in the next sections we will pursue deeper investigations of three representative cases from the set we have examined here: one in which the vortex is centered at the root, and two (of opposite sign) in which the vortex is centered at the left tip. With these, we seek to reveal the manner in which the observed lobed vortical structures form and evolve, and how they contribute to the aerodynamic loads.

\subsubsection{Surface signatures and flow kinematics}\label{subsec:surface_signatures}

\begin{figure}[htbp]
    \centering
    \begin{overpic}[width=1.0\linewidth]{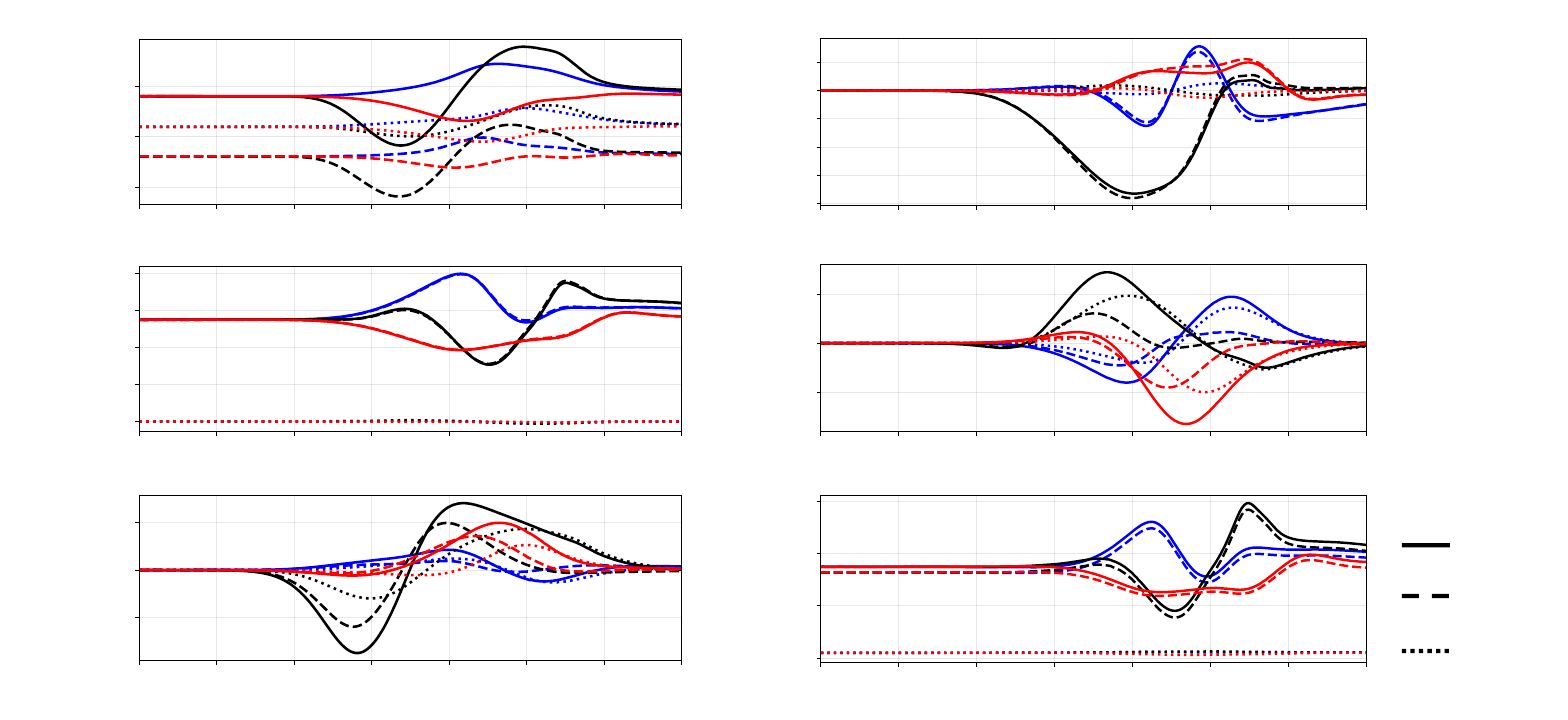}
    \put(6.6,1.8){\small{$-1.5$}}
    \put(11.6,1.8){\small{$-1.0$}}
    \put(16.6,1.8){\small{$-0.5$}}
    \put(22.6,1.8){\small{$0.0$}}
    \put(27.6,1.8){\small{$0.5$}}
    \put(32.6,1.8){\small{$1.0$}}
    \put(37.6,1.8){\small{$1.5$}}
    \put(42.6,1.8){\small{$2.0$}}

    \put(6.6,16.5){\small{$-1.5$}}
    \put(11.6,16.5){\small{$-1.0$}}
    \put(16.6,16.5){\small{$-0.5$}}
    \put(22.6,16.5){\small{$0.0$}}
    \put(27.6,16.5){\small{$0.5$}}
    \put(32.6,16.5){\small{$1.0$}}
    \put(37.6,16.5){\small{$1.5$}}
    \put(42.6,16.5){\small{$2.0$}}

    \put(6.6,31){\small{$-1.5$}}
    \put(11.6,31){\small{$-1.0$}}
    \put(16.6,31){\small{$-0.5$}}
    \put(22.6,31){\small{$0.0$}}
    \put(27.6,31){\small{$0.5$}}
    \put(32.6,31){\small{$1.0$}}
    \put(37.6,31){\small{$1.5$}}
    \put(42.6,31){\small{$2.0$}}

    \put(50.2,1.7){\small{$-1.5$}}
    \put(55.2,1.7){\small{$-1.0$}}
    \put(60.2,1.7){\small{$-0.5$}}
    \put(66.2,1.7){\small{$0.0$}}
    \put(71.2,1.7){\small{$0.5$}}
    \put(76.2,1.7){\small{$1.0$}}
    \put(81.2,1.7){\small{$1.5$}}
    \put(86.2,1.7){\small{$2.0$}}

    \put(50.2,16.5){\small{$-1.5$}}
    \put(55.2,16.5){\small{$-1.0$}}
    \put(60.2,16.5){\small{$-0.5$}}
    \put(66.2,16.5){\small{$0.0$}}
    \put(71.2,16.5){\small{$0.5$}}
    \put(76.2,16.5){\small{$1.0$}}
    \put(81.2,16.5){\small{$1.5$}}
    \put(86.2,16.5){\small{$2.0$}}

    \put(50.2,31){\small{$-1.5$}}
    \put(55.2,31){\small{$-1.0$}}
    \put(60.2,31){\small{$-0.5$}}
    \put(66.2,31){\small{$0.0$}}
    \put(71.2,31){\small{$0.5$}}
    \put(76.2,31){\small{$1.0$}}
    \put(81.2,31){\small{$1.5$}}
    \put(86.2,31){\small{$2.0$}}

    \put(3.2,5.9){\small{$-0.05$}}
    \put(4.7,8.9){\small{$0.00$}}
    \put(4.7,11.9){\small{$0.05$}}

    \put(5.7,18.5){\small{$0.0$}}
    \put(5.7,20.8){\small{$0.1$}}
    \put(5.7,23.1){\small{$0.2$}}
    \put(5.7,25.4){\small{$0.3$}}
    \put(5.7,27.7){\small{$0.4$}}

    \put(5.7,33.3){\small{$0.0$}}
    \put(5.7,36.6){\small{$0.1$}}
    \put(5.7,39.9){\small{$0.2$}}

    \put(48.3,3.5){\small{$0.00$}}
    \put(48.3,6.7){\small{$0.05$}}
    \put(48.3,9.9){\small{$0.10$}}
    \put(48.3,13.1){\small{$0.15$}}

    \put(46.8,20.3){\small{$-0.02$}}
    \put(48.3,23.4){\small{$0.00$}}
    \put(48.3,26.5){\small{$0.02$}}

    \put(46.8,32.4){\small{$-0.04$}}
    \put(46.8,34.1){\small{$-0.03$}}
    \put(46.8,35.8){\small{$-0.02$}}
    \put(46.8,37.5){\small{$-0.01$}}
    \put(48.3,39.4){\small{$0.00$}}
    \put(48.3,41.1){\small{$0.01$}}

    \put(1,36.5){\rotatebox{90}{\large$C_D$}}
    \put(1,22.5){\rotatebox{90}{\large$C_L$}}
    \put(1,8){\rotatebox{90}{\large$C_S$}}

    \put(44.9,36.5){\rotatebox{90}{\large$C_{M_r}$}}
    \put(44.9,22){\rotatebox{90}{\large$C_{M_y}$}}
    \put(44.9,7.5){\rotatebox{90}{\large$C_{M_p}$}}

    \put(25.5,0){{\large$t$}}
    \put(69,0){{\large$t$}}

    \put(93.2,10.6){{total}}
    \put(93.2,7.4){{pressure}}
    \put(93.2,3.9){{viscous}}
    \end{overpic}
    \caption{the aerodynamic force and moment histories during gust encounters with $(\phi,z_0)=(\pi/2,0)$ (black), $(\phi,z_0)=(\pi/2,0.6)$ (red), and $(\phi,z_0)=(3\pi/2,0.6)$ (blue).}
    \label{fig:aero_z0_3}
\end{figure}

The flow dynamics and force and moment responses are connected through the surface signatures imprinted by the gust on the delta wing, including the surface pressure and skin-friction line patterns, and we examine those patterns in this section and connect them with the flow kinematics. Before doing so, in figure \ref{fig:aero_z0_3} we decompose the aerodynamic response of our three representative cases into the pressure contribution and the viscous contribution, to clarify the fluid dynamic origins of the force and moment. We can observe that the pressure influence dominates the responses of lift force and the roll and pitching moments, while the contributions are comparably important in the drag and side force and yaw moment responses. We will refer frequently to this figure during our forthcoming analysis of the surface pressure and skin friction.

\begin{figure}[htbp]
    \centering
    \begin{overpic}[width=1.0\linewidth]{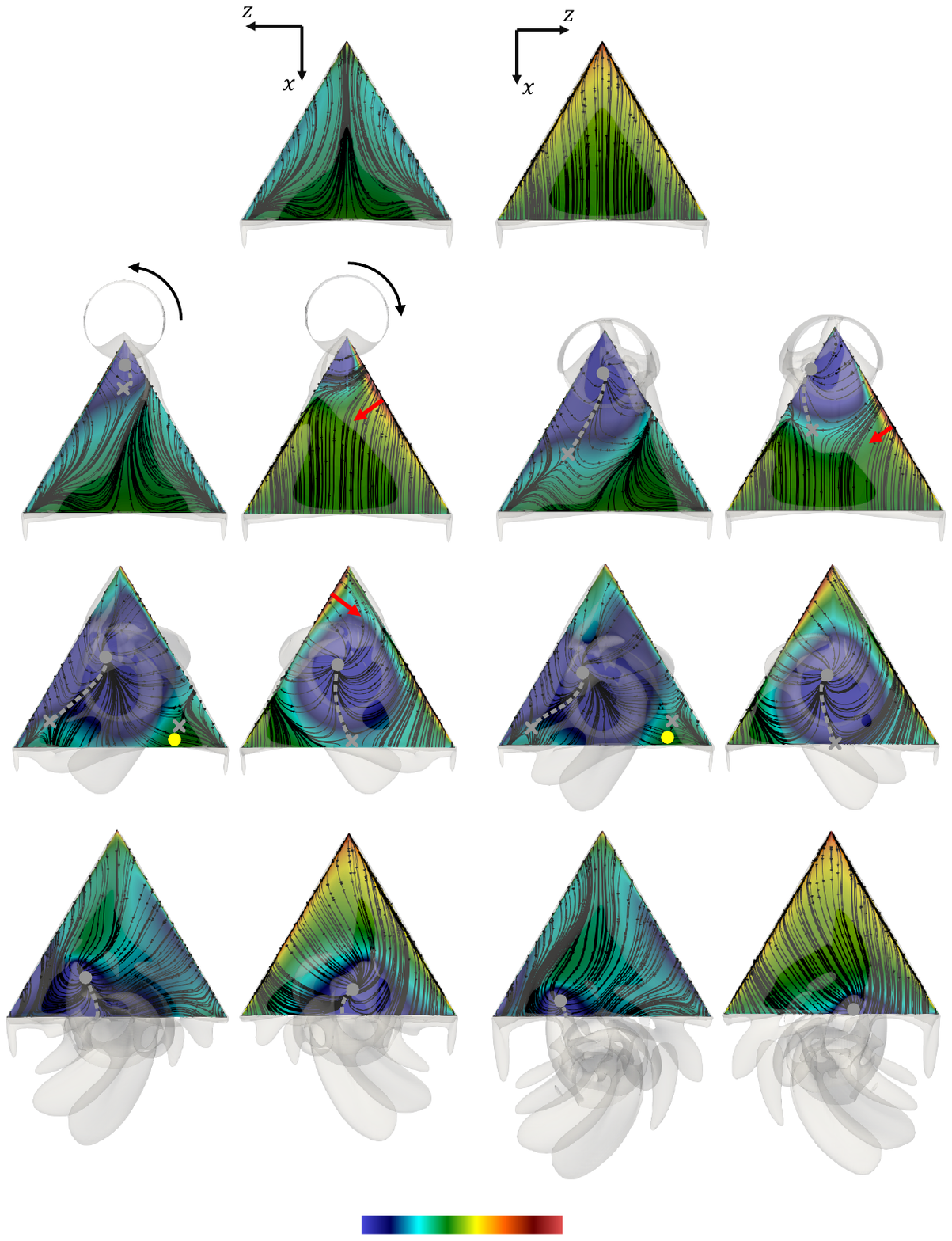}
    \put(15,100){{$(a)$}}
    \put(0,73){{$(b)$}}
    \put(38,73){{$(c)$}}
    \put(26.4,80.5){{(upper)}}
    \put(46.7,80.5){{(lower)}}
    \put(0,53.5){{$(d)$}}
    \put(38,53.5){{$(e)$}}
    \put(0,32){{$(f)$}}
    \put(38,32){{$(g)$}}
    \put(36.8,-0.3){{$C_P$}}
    \put(28.5,3.2){{$-1$}}
    \put(45.4,3.2){{$1$}}
    \end{overpic}
    \caption{The skin-friction line patterns and the surface pressure contours on both the upper and lower sides, overlaid by instantaneous $Q$ isosurface, for gust encounter $(\phi, z_0)=(\pi/2,0)$ at $(a)~t=-1.5$, $(b)~t=-0.1$, $(c)~t=0.2$, $(d)~t=0.6$, $(e)~t=0.7$, $(f)~t=1.0$, and $(g)~t=1.2$. Grey circle indicates the induced nodal point of separation, and yellow circle indicates the induced nodal point of attachment. Grey cross represents the saddle point, and the dashed line represents the separation line.}
    \label{fig:surface_midspan}
\end{figure}

In figure \ref{fig:surface_midspan} we show the surface plots of pressure and shear stress on the upper and lower wing surfaces for the encounter of a counter-clockwise gust with the root (i.e., case ID $2$). As we can see in panel (a), the pressure distribution and the skin-friction line pattern are symmetric about the root chord in the absence of the gust. On the upper surface, the region just aft of the leading edge is observed to have a pressure drop, which is then followed by an adverse pressure gradient until the trailing edge. The adverse pressure gradient is moderate, so we do not observe separation; skin friction lines emerge from a single attachment node at the apex and are consistent the viscous contribution to drag. However, the skin-friction lines starting from the leading edge tend to converge at the wing tips and exhibit some flow reversal near the trailing edge, so we consider this baseline flow to be on the verge of separation. The flow on the lower surface is fully attached.

As the vortex approaches and passes over the wing, it has three important effects that can be observed in these surface patterns. Most obvious of these is the traction exerted on the upper and lower surfaces, evident in the lower-pressure core observed on both surfaces. These cores nearly negate each other's contribution to normal force, as evident in the negligible disturbance in lift before $t = 0.3$ in figure~\ref{fig:aero_z0_3}. However, this balance is disrupted in the later stages, first shifting in favor of the lower side in (e), decreasing the lift and pitching moment, and then shifting to the upper side in (f) and (g) as the induced lobed structures imprint on the upper surface pressure, thereby increasing the lift and pitching moment. A slight imbalance in the left-right symmetry of this balance creates a small rolling moment about the negative $x$ axis for the duration of the encounter. The lateral re-orientation of the lines of skin friction is consistent with the trends in the viscous components of side force and yaw moment in figure~\ref{fig:aero_z0_3}.

The vortex also has a strong influence on pressure along the leading edge, and to clarify this we will use the terms `leeward' and `windward' to describe the sides of the leading edge with respect to the flow induced by the vortex gust. Early in the encounter, the vortex generates suction on the leeward side near the apex, as we can see from figure \ref{fig:surface_midspan}. This suction is responsible for the early decrease in drag in figure~\ref{fig:aero_z0_3} and reaches its peak influence in panel (c). On the windward side further aft, the vortex induces stagnation, as indicated by the red arrows, most apparent in the elevated pressure near the leading edge on the lower surface. The stagnation and suction cooperatively generate the disturbances in the pressure contributions to side force and yaw moment in figure~\ref{fig:aero_z0_3}. In panel (d), with the vortex near the trailing edge and the leeward and windward sides now swapped behind the vortex, the side force peaks in the opposite direction. The center of this side force is near the quarter chord, evident from the nearly vanishing pressure contribution to yaw moment. 

Finally, the low-pressure core creates regions of adverse pressure gradient that affect the boundary layers on both the upper and lower surfaces, as revealed by the topology of the lines of skin friction in figure~\ref{fig:surface_midspan}. Rather than labeling and describing every critical point, we focus only on lines connecting saddle points and attracting nodes, along which separation is expected if there is sufficient rate of attraction \cite{Lighthill1963}; these are labeled in figure~\ref{fig:surface_midspan}. The first such separation line appears on the upper surface in panel (b) and on the lower surface in panel (c); these lines elongate and move aft on both surfaces before they disappear off the trailing edge. The separation lines on both surfaces are associated with a lifting of the boundary layer from the surface, starting at the upper surface near the apex. 

\begin{figure}[htbp]
    \centering
    \begin{overpic}[width=1.0\linewidth]{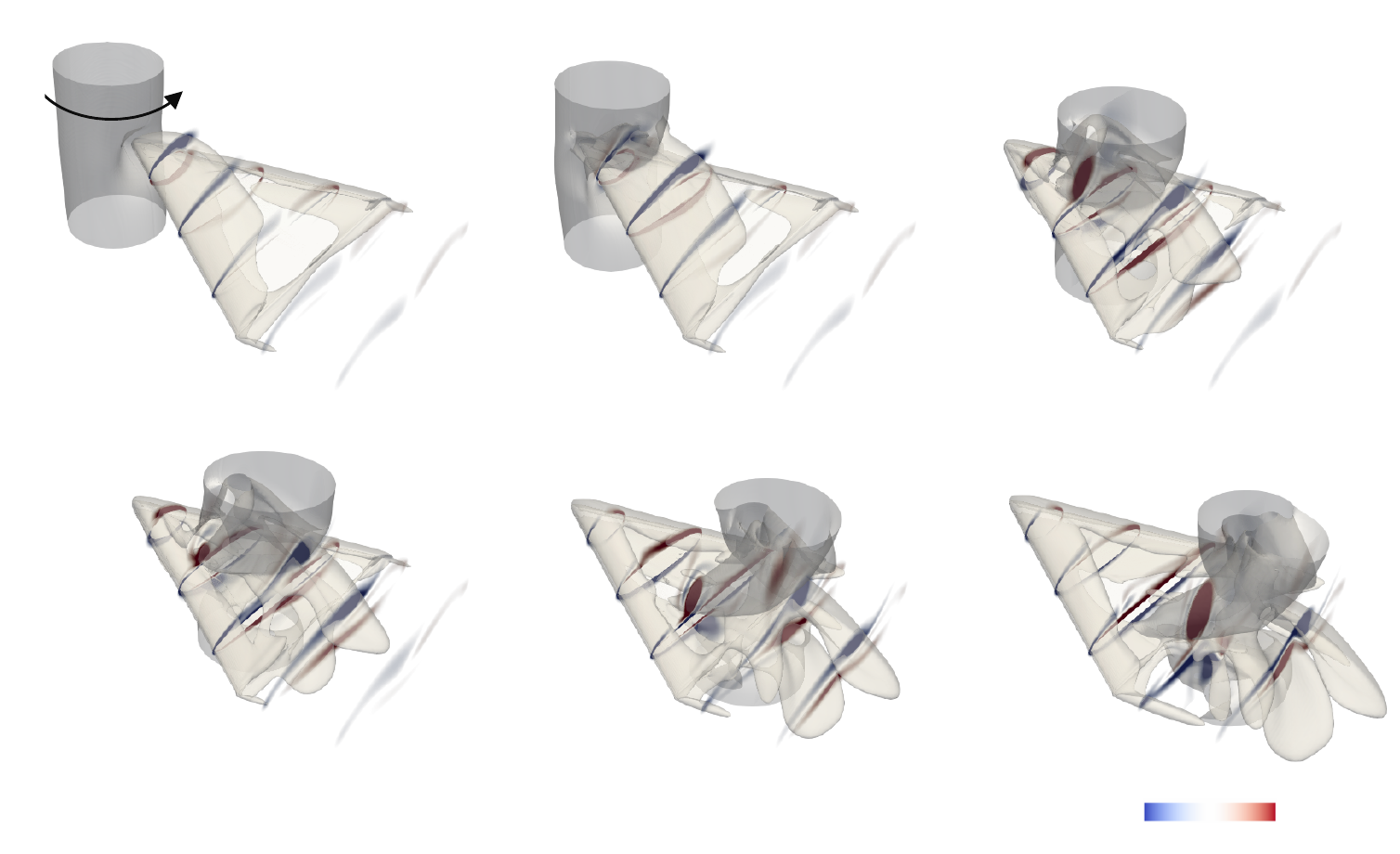}
    \put(0,60){{$(a)$}}
    \put(34,60){{$(b)$}}
    \put(68,60){{$(c)$}}
    \put(0,30){{$(d)$}}
    \put(34,30){{$(e)$}}
    \put(68,30){{$(f)$}}
    \put(85.5,0.5){{$\omega_x$}}
    \put(79.7,4.3){{\small$-10$}}
    \put(90,4.3){{\small$10$}}
    \put(16.5,45){{$1$}}
    \put(14.5,43){{$2$}}
    \put(51.5,44){{$1$}}
    \put(48.5,42){{$2$}}
    \put(85,43.5){{$1$}}
    \put(83,38.5){{$2$}}
    \put(75.8,47.1){{$3$}}
    \put(26,15){{$1$}}
    \put(22,12.1){{$2$}}
    \put(15.1,21){{$3$}}
    \put(62,12.5){{$1$}}
    \put(57.5,10){{$2$}}
    \put(50.5,18){{$3$}}
    \put(96,12){{$1$}}
    \put(92.5,8.5){{$2$}}
    \put(83,16.1){{$3$}}
    \end{overpic}
    \caption{Gust encounter $(\phi,z_0)=(\pi/2,0)$: the streamwise vorticity $\omega_x$ contours on transverse planes at $x=0.2$, $x=0.5$, $x=0.8$, $x=1.1$, and $x=1.4$, overlaid by the instantaneous $Q=3.5$ isosurface, at (a)~$t=-0.1$, (b)$~t=0.2$, (c)~$t=0.6$, (d)~$t=0.7$, (e)~$t=1.0$, and (f)~$t=1.2$.}
    \label{fig:vort_z_0_p}
\end{figure}
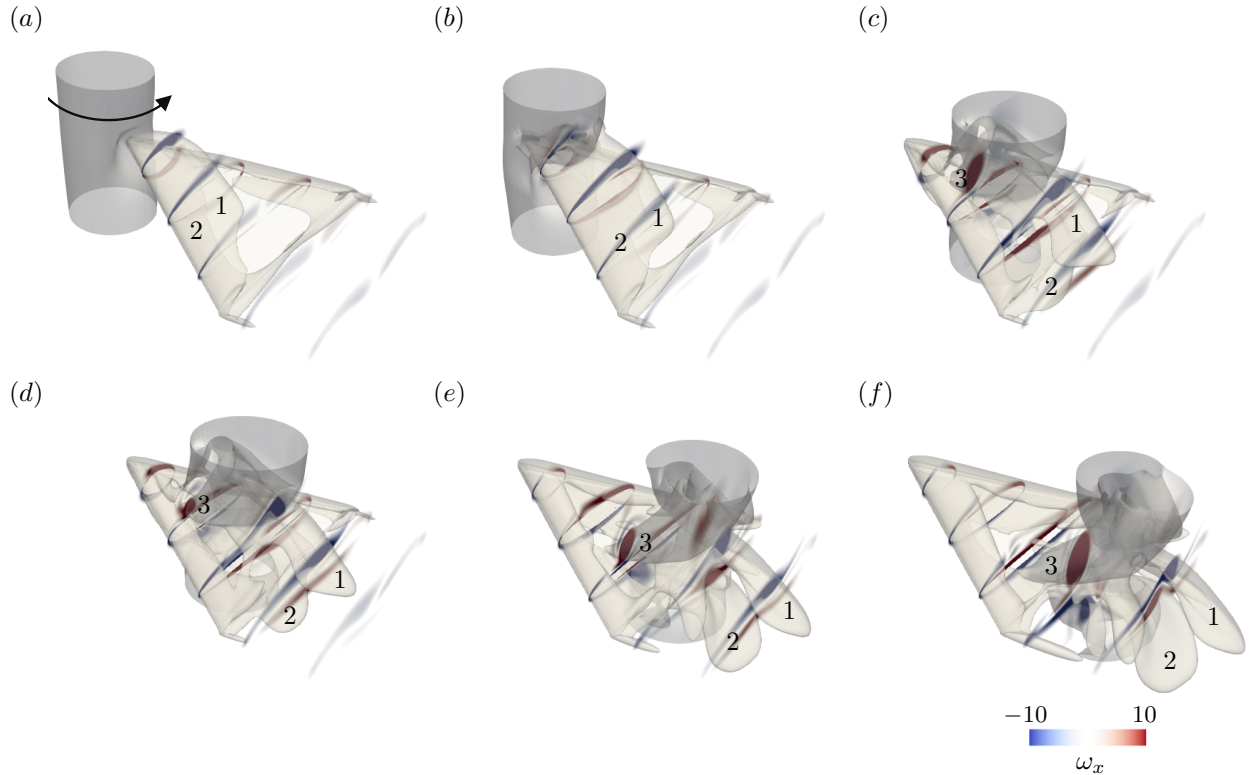

To understand what happens to this lifted boundary layer, figure \ref{fig:vort_z_0_p} shows a $Q$ isosurface along with streamwise vorticity $\omega_x$ on several transverse slices, at several instants during the gust encounter. In the early stages when the gust is near the apex of the wing, as shown in panel (a) and (b), the lifted boundary layer on the windward upper surface is pulled across the wing by the cross-flow of the gust, and slightly later, the boundary layer on the windward lower surface experiences a similar treatment. Later, in (c) and (d), when the gust has traveled halfway down the wing, these lifted boundary layers form a pair of lobed vortical structures (labeled as $1$ and $2$) with opposite sign vorticity. Meanwhile, the boundary layer on the leeward upper side near the apex, lifted along the separation line that emerged near the leading edge of the lower side (panel (c) in figure~\ref{fig:surface_midspan}), is collected and strained by the opposing flows from the vortex gust and freestream to form another lobe (labeled as $3$) behind the advancing gust. All three lobes are twisted around the gust core as it leaves the wing, in panels (e) and (f).

\begin{figure}[htbp]
    \centering
    \begin{overpic}[width=1.0\linewidth]{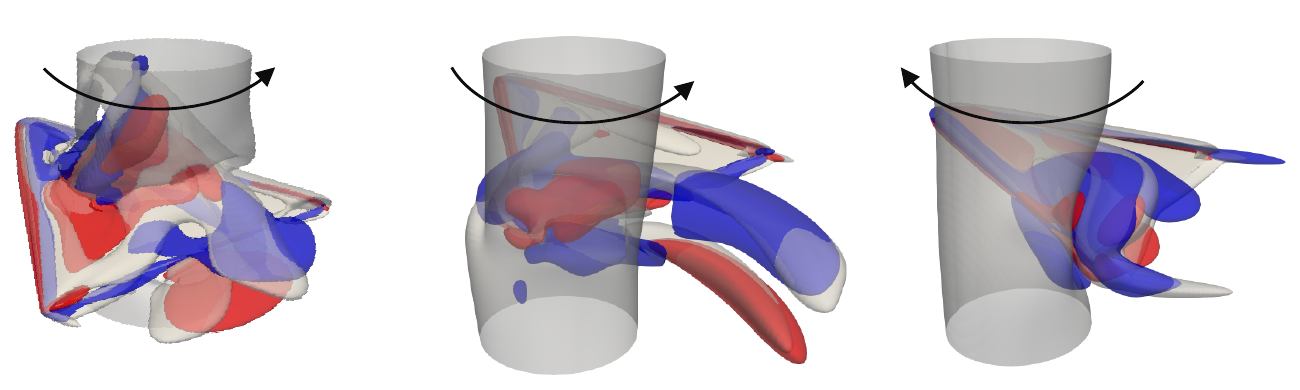}
    \put(0,30){{$(a)$}}
    \put(34,30){{$(b)$}}
    \put(68,30){{$(c)$}}
    \put(19,10){{$1$}}
    \put(14,6.5){{$2$}}
    \put(6,17){{$3$}}
    \put(61.9,10){{$1$}}
    \put(58,6.5){{$2$}}
    \put(84.9,12.5){{$1$}}
    \put(86.7,10.1){{$2$}}
    \end{overpic}
    \caption{The instantaneous isosurfaces of $Q$ (grey color) and selected component of $\boldsymbol T$ (red color for positive value and blue color for negative value) in each gust encounter: $(a)$ $T_x$ for case $(\phi,z_0)=(\pi/2,0)$, $(b)$ $T_x$ for case $(\phi,z_0)=(\pi/2,0.6)$, and $(c)$ $T_z$ for case $(\phi,z_0)=(3\pi/2,0.6)$.}
    \label{fig:stretch}
\end{figure}

While the above observations reveal the geometric evolution of the lobed vortical structures, the mechanism responsible for that evolution remains to be clarified. Therefore, in figure~\ref{fig:stretch}(a) we inspect the stretching and tilting term, $\boldsymbol{T} = \boldsymbol{\omega} \cdot \nabla \boldsymbol{u}$, responsible for processing each component of vorticity by the local strain field. In particular, if a particular component $T_i$ has the same sign as $\omega_i$, that vorticity component is intensified, whereas an opposite sign indicates a weakening. In figure \ref{fig:vort_z_0_p} we used the $x$ (streamwise) component of vorticity to reveal the creation of the lobes in this gust encounter. Therefore, we use $T_x$ as a representative component along with an isosurface of $Q$ to illuminate the correspondence between vortex stretching/tilting and the geometric evolution of the vortical structures in figure~\ref{fig:stretch}(a). As we can see, lobe $1$, which has negative streamwise vorticity $\omega_x$ in figure \ref{fig:vort_z_0_p}, shows a fair agreement with the negative isosurface of $T_x$, indicating that the streamwise vorticity within this structure is not merely advected but locally intensified by the gust strain field. A similar intensification is observed for lobe 2, which has a positive streamwise vorticity $\omega_x$ and aligns with the positive isosurface of $T_x$. In contrast, lobe $3$ shows a more complex signature containing adjacent regions of both positive and negative $T_x$ in a lobe of overall positive streamwise vorticity. This can be attributed to the highly non-uniform strain field at the junction of the gust and the wing surface. Therefore, this lobe's processing varies between favorable stretching and attenuating compression. 

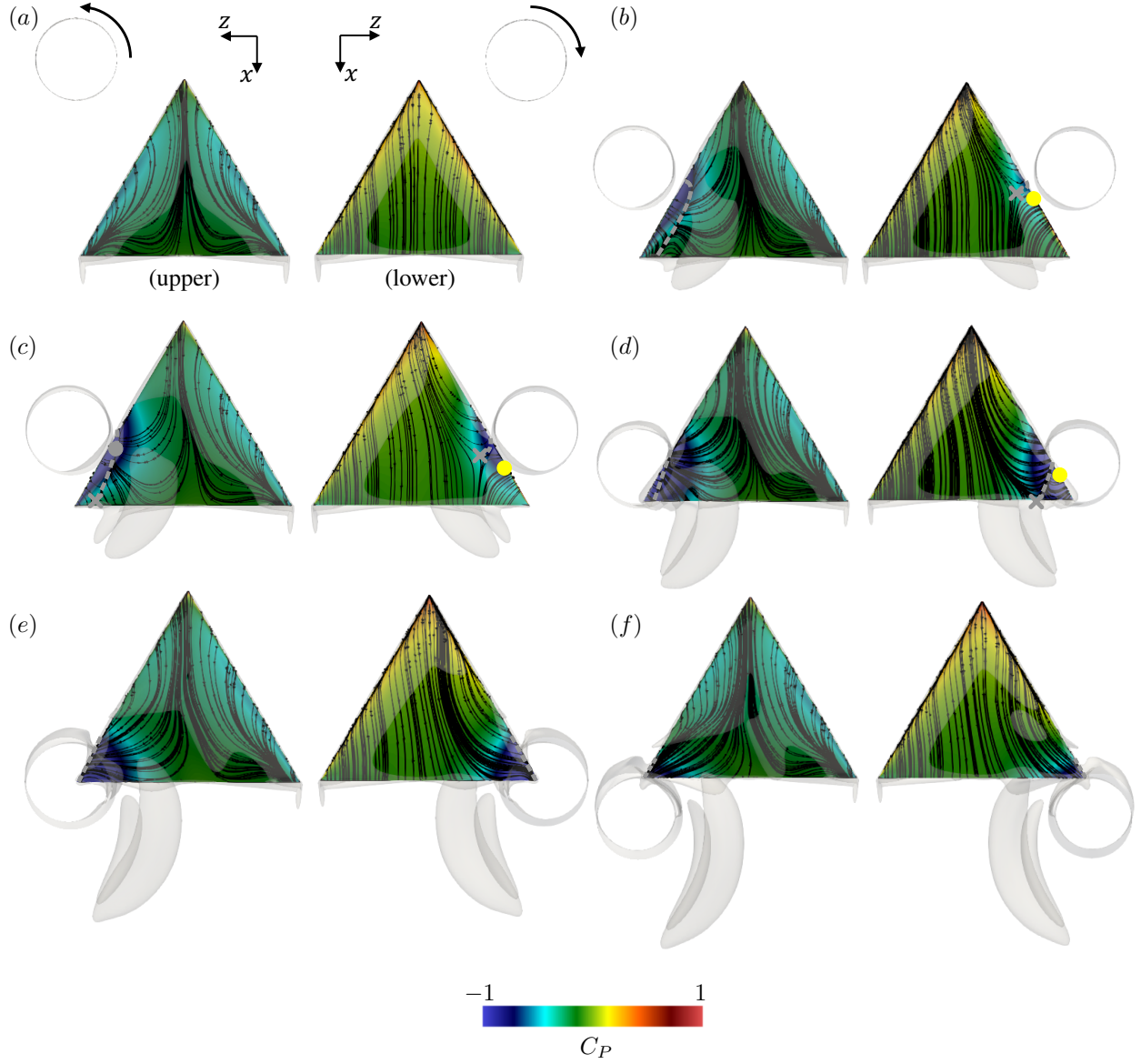
\begin{figure}[htbp]
    \centering
    \begin{overpic}[width=1.0\linewidth]{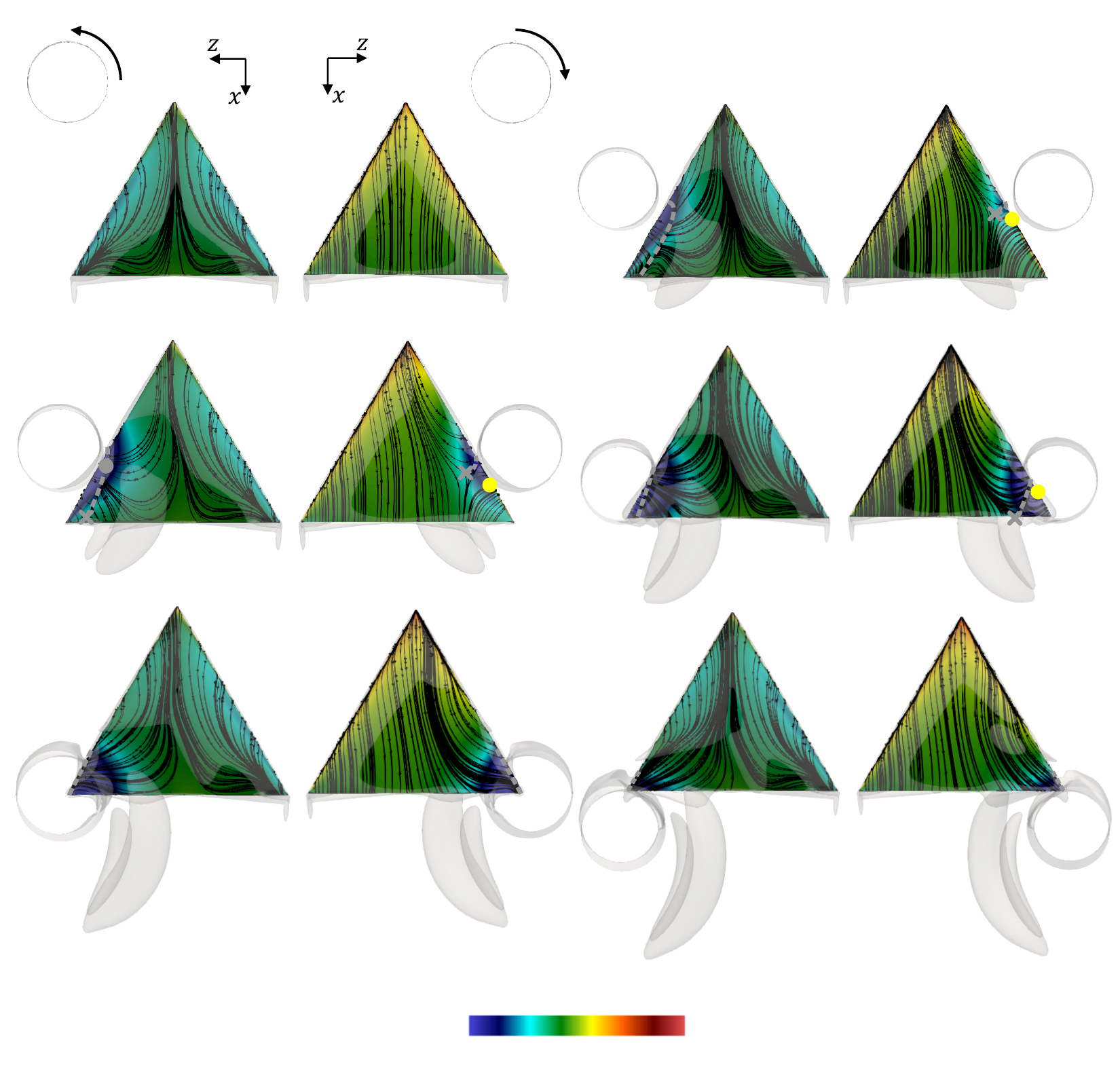}
    \put(0,91){{$(a)$}}
    \put(53,91){{$(b)$}}
    \put(0,62){{$(c)$}}
    \put(12.4,68){{(upper)}}
    \put(33.2,68){{(lower)}}
    \put(53,62){{$(d)$}}
    \put(0,37.5){{$(e)$}}
    \put(53,37.5){{$(f)$}}
    \put(50.4,0.3){{$C_P$}}
    \put(40.2,5){{$-1$}}
    \put(60.5,5){{$1$}}
    \end{overpic}
    \caption{The skin-friction line patterns and the surface pressure contours on both the upper and lower sides, overlaid by instantaneous $Q$ isosurface, for gust encounter $(\phi, z_0)=(\pi/2,0.60)$ at $(a)~t=-0.1$, $(b)~t=0.5$, $(c)~t=0.6$, $(d)~t=0.8$, $(e)~t=1.0$, and $(f)~t=1.2$. Grey circle indicates the induced nodal point of separation, and yellow circle indicates the induced nodal point of attachment. Grey cross represents the saddle point, and the dashed line represents the separation line.}
    \label{fig:surface_z_0d6}
\end{figure}

Having established the physical process for the centered counter-clockwise gust encounter, we now examine how the process is modified when the encounter is shifted laterally toward the left tip (i.e., case ID $8$). The surface signatures in this case are shown in figure \ref{fig:surface_z_0d6}. The disturbance is only apparent after the gust has traveled half of the chord, in panel (b), when the low-pressure core imprints on the left side of the wing. The associated suction exerted on the left leading edge only slightly decreases the drag in figure~\ref{fig:aero_z0_3}, but creates a stronger disruption of the side force and yaw moment. A saddle point appears on the lower side of the wing, indicating that flow on this side is redirected from the apex around the leading edge and separates along a line on the left side of the upper surface, lifting the boundary layer off this surface to form, with the original tip vortex, an extended lobed structure. By panel (c), a new attracting node and saddle point appear on the upper surface, and the original saddle on the lower surface has migrated toward the trailing edge. This creates a new separation line on the lower surface that lifts the boundary layer from that surface to form a second lobe.

\begin{figure}[htbp]
    \centering
    \begin{overpic}[width=1.0\linewidth]{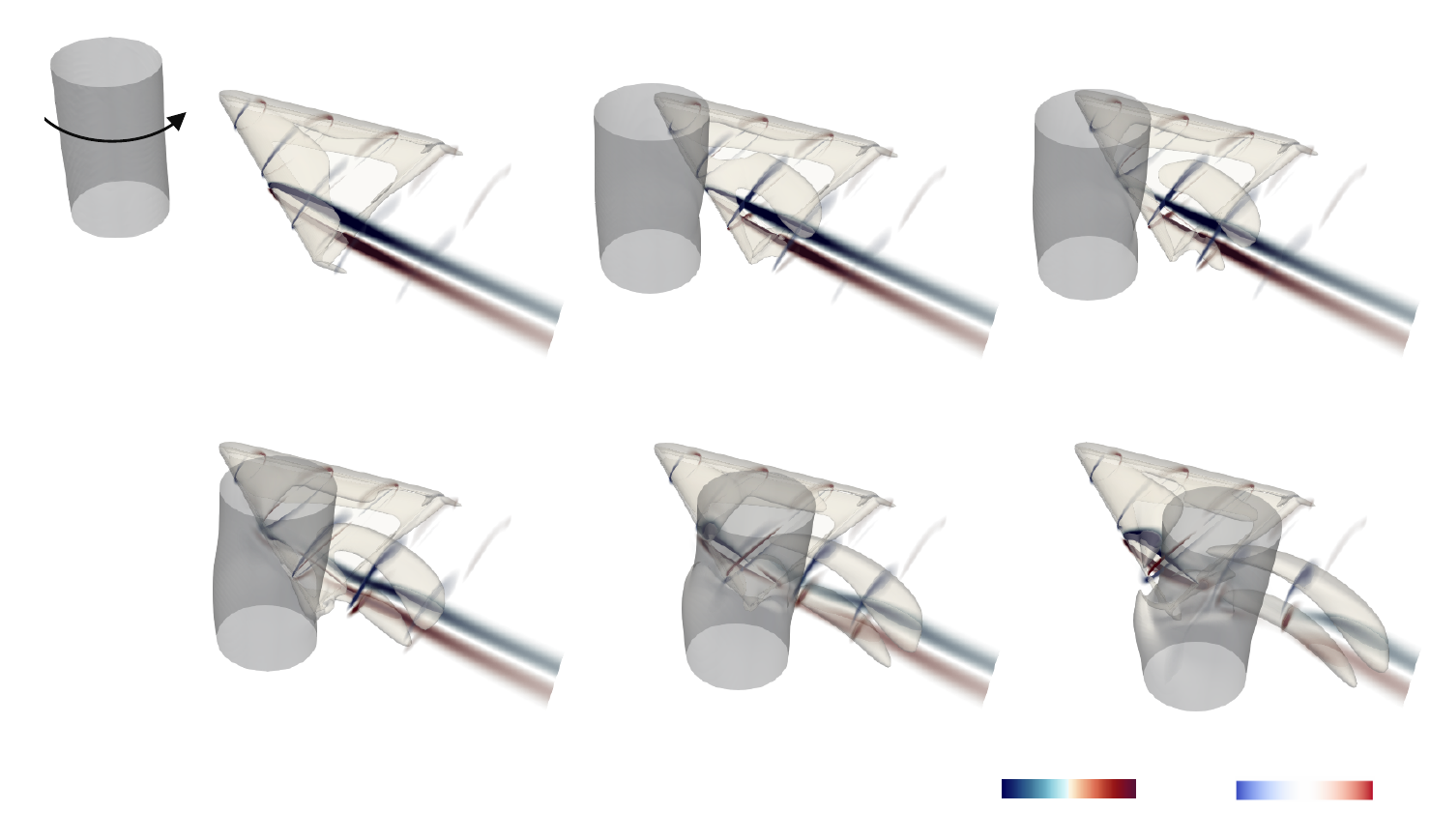}
    \put(1,57){{$(a)$}}
    \put(40,57){{$(b)$}}
    \put(71,57){{$(c)$}}
    \put(1,31){{$(d)$}}
    \put(40,31){{$(e)$}}
    \put(71,31){{$(f)$}}
    \put(90.2,1){{$\omega_x$}}
    \put(84.5,4.7){{\small$-10$}}
    \put(94.8,4.7){{\small$10$}}
    \put(73.8,1){{$\omega_z$}}
    \put(68,4.7){{\small$-10$}}
    \put(78.3,4.7){{\small$10$}}
    \put(21.1,47){{$1$}}
    \put(23,40.7){{$2$}}
    \put(54.9,44.5){{$1$}}
    \put(53.9,40.7){{$2$}}
    \put(84.7,44.5){{$1$}}
    \put(84.1,40.7){{$2$}}
    \put(28.7,17.4){{$1$}}
    \put(26.7,14.9){{$2$}}
    \put(61.7,16.5){{$1$}}
    \put(58.6,15.1){{$2$}}
    \put(93.7,15.7){{$1$}}
    \put(90.6,14.2){{$2$}}
    \end{overpic}
    \caption{Gust encounter $(\phi,z_0)=(\pi/2,0.60)$: the streamwise vorticity $\omega_x$ contours on transverse planes at $x=0.2$, $x=0.5$, $x=0.8$, $x=1.1$, and $x=1.4$, overlaid by the instantaneous $Q=3.5$ isosurface, at (a)~$t=-0.1$, (b)~$t=0.5$, (c)~$t=0.6$, (d)~$t=0.8$, (e)~$t=1.0$, and (f)~$t=1.2$.}
    \label{fig:vort_z_0d6_p}
\end{figure}

As we did for the centered gust encounter, we use $Q$ isosurfaces and contours of $\omega_x$ in transverse planes, shown in Figure \ref{fig:vort_z_0d6_p}, to clarify the kinematics of the lobes. Here, we also show the spanwise vorticity $\omega_z$ on a representative streamwise plane at $z=0.35$. Lobes $1$ and $2$ are clearly revealed by panels (b) and (c), respectively, where they have advected inboard. The streamwise vorticity $\omega_x$ at slices of $x=0.8$ and $x=1.1$ becomes stronger, while the spanwise vorticity $\omega_z$ at slice of $z=0.35$ is weakened near the leading edge due to the opposing flows in this region. These lobes are elongated and slightly wrapped around the vortex gust, as shown in panels (d) to (f). In figure~\ref{fig:stretch}(b) we use $T_x$ as a representative component of the stretching/tilting with the $Q$ isosurface to illuminate the influence from the gust's strain field. Here, we find a good alignment between the isosurfaces of $T_x$ and like-signed vorticity $\omega_x$ in both lobes, indicating an intensification of vorticity in each.


\begin{figure}[htbp]
    \centering
    \begin{overpic}[width=1.0\linewidth]{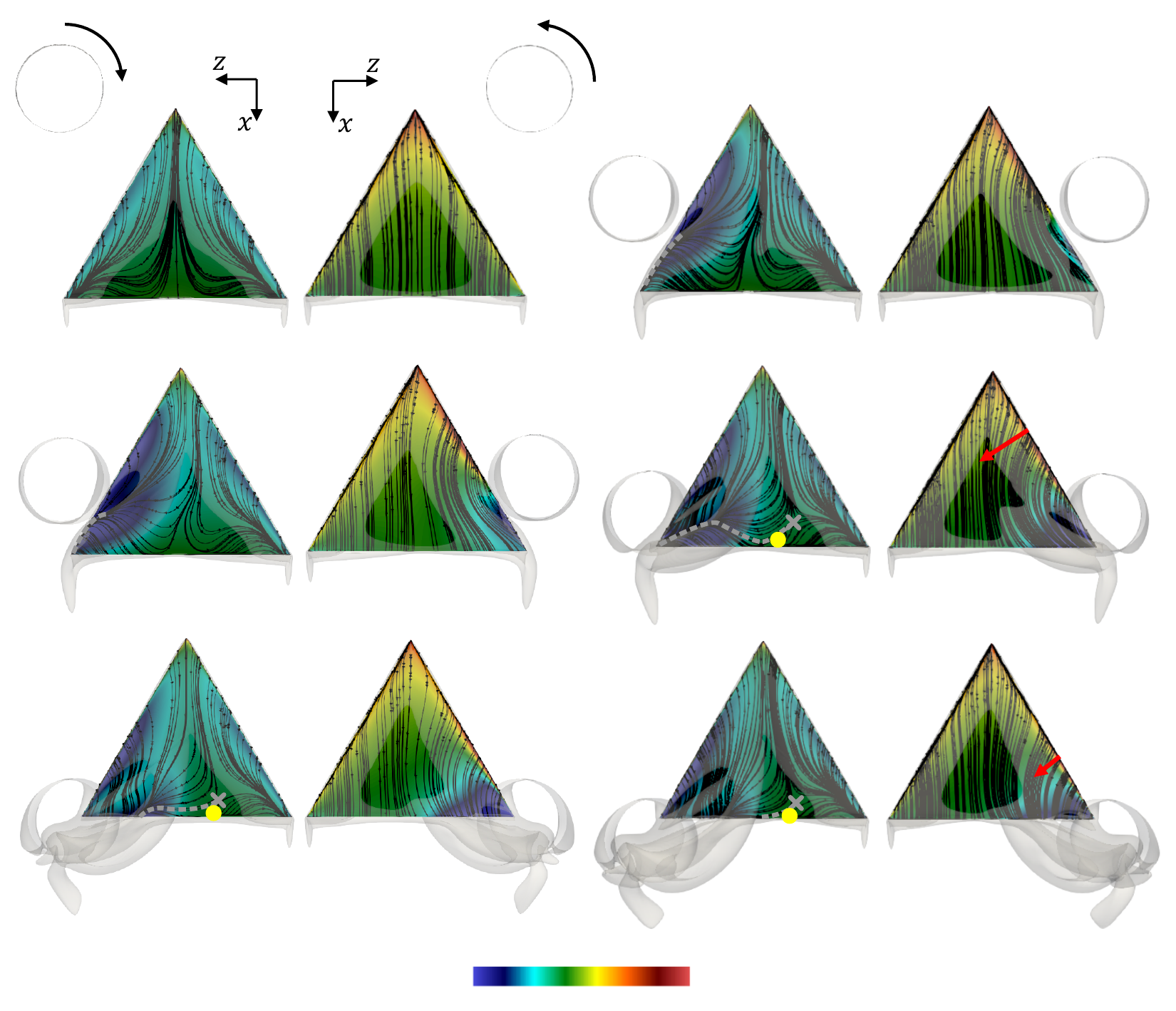}
    \put(0,84.5){{$(a)$}}
    \put(52,84.5){{$(b)$}}
    \put(0,53){{$(c)$}}
    \put(11.7,58.5){{(upper)}}
    \put(32.4,58.5){{(lower)}}
    \put(52,53){{$(d)$}}
    \put(0,29){{$(e)$}}
    \put(52,29){{$(f)$}}
    \put(48.3,0){{$C_P$}}
    \put(38.6,4.3){{$-1$}}
    \put(58.1,4.3){{$1$}}
    \end{overpic}
    \caption{The skin-friction line patterns and the surface pressure contours on both the upper and lower sides, overlaid by instantaneous $Q$ isosurface, for gust encounter $(\phi, z_0)=(3\pi/2,0.60)$ at $(a)~t=-0.1$, $(b)~t=0.5$, $(c)~t=0.6$, $(d)~t=0.8$, $(e)~t=1.0$, and $(f)~t=1.1$. Yellow circle indicates the induced spiral point of attachment. Grey cross represents the saddle point, and the dashed line represents the open separation line.}
    \label{fig:surface_z_0d6_n}
\end{figure}

We contrast this counter-clockwise vortex encounter at the left tip with a clockwise one (case ID $12$), for which the surface signatures are shown in figure \ref{fig:surface_z_0d6_n}. As in the counter-clockwise case, the clockwise gust begins to imprint its low-pressure core on the left edge once it convects to mid-chord, as indicated in panel (b). However, there are important differences that arise from the fact that the vortex's induced flow in this vicinity is in the same direction as the free stream. The most obvious is apparent on the upper surface, where the baseline streamwise flow is strongest and the low-pressure region consequently larger than in the clockwise case. This suction has the effect of increasing the lift force more substantially than the other two cases. The coordinated flows from the gust and free stream also create a stagnation region on the leading edge behind the advancing vortex. This edge pressure (indicated with a red arrow) increases the drag, but its influence on the side force and yaw moment is offset by the edge suction exerted further aft.

Behind this low-pressure core is an adverse pressure gradient that induces a region of flow reversal near the left leading edge, demarcated by an open separation line (a line not connected to a saddle point \cite{Wang1972,Wang1976,Tobak1982,Yates1992}). The separated flow lifts the boundary layer from the left side of the upper surface and enhances the adjacent tip vortex.  By panel (d), the open separation line, now connected with a new saddle point and therefore closed, moves aft toward the trailing edge. The lifted boundary layer is swept into the wake and forms an extended lobe structure with the tip vortex in panels (e) and (f).

\begin{figure}[htbp]
    \centering
    \begin{overpic}[width=1.0\linewidth]{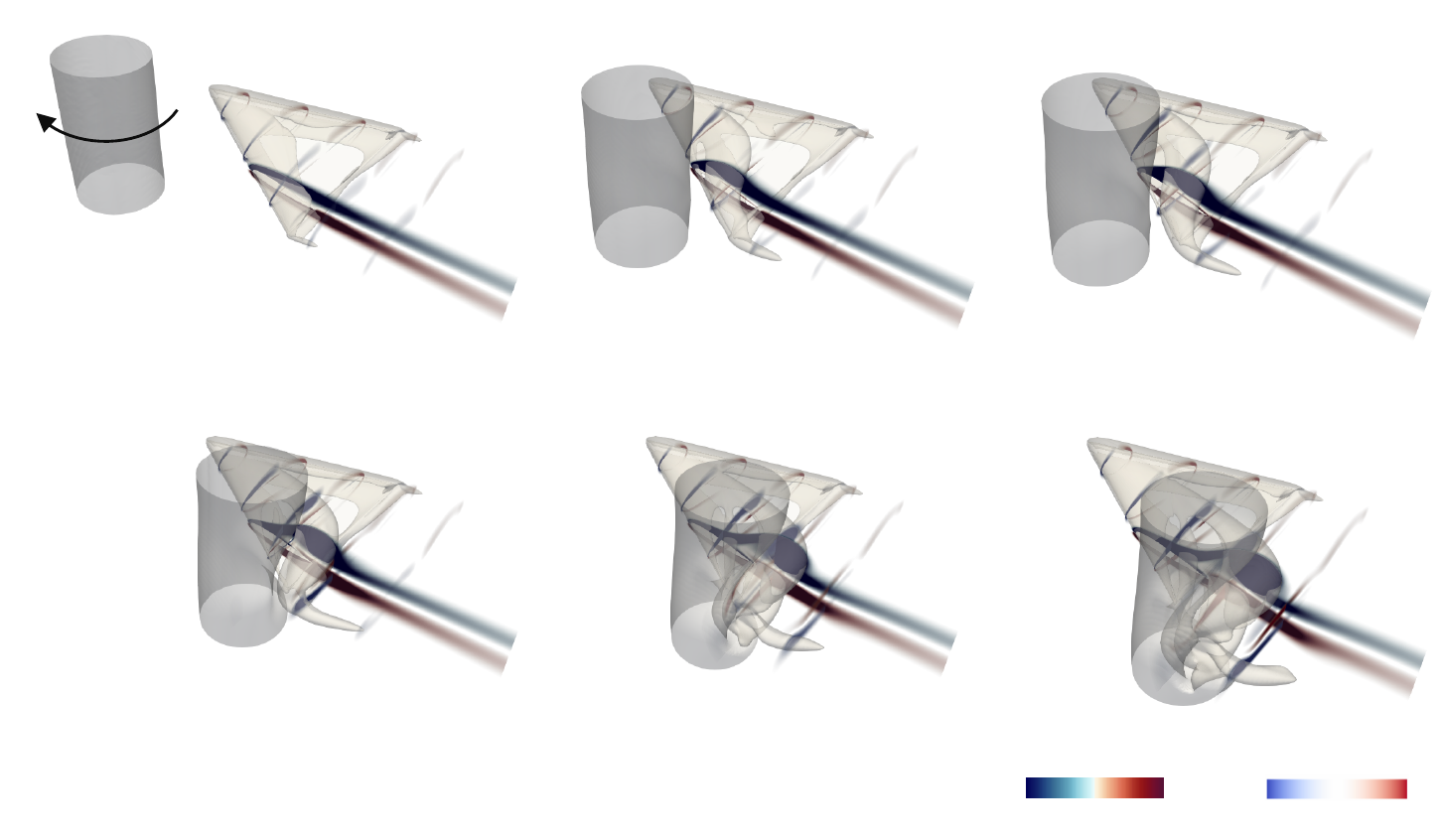}
    \put(0,56.5){{$(a)$}}
    \put(36,56.5){{$(b)$}}
    \put(68,56.5){{$(c)$}}
    \put(0,31){{$(d)$}}
    \put(36,31){{$(e)$}}
    \put(68,31){{$(f)$}}
    \put(91.7,1){{$\omega_x$}}
    \put(85.8,4.7){{\small$-10$}}
    \put(96.3,4.7){{\small$10$}}
    \put(75,1){{$\omega_z$}}
    \put(69,4.7){{\small$-10$}}
    \put(79.5,4.7){{\small$10$}}
    \put(19,47.5){{$1$}}
    \put(17.3,47){{$2$}}
    \put(50.7,47.5){{$1$}}
    \put(49.2,47){{$2$}}
    \put(50.7,47.5){{$1$}}
    \put(49.2,47){{$2$}}
    \put(82.2,47.5){{$1$}}
    \put(80.9,46.7){{$2$}}
    \put(21.6,20.4){{$1$}}
    \put(20.3,18.7){{$2$}}
    \put(54,19){{$1$}}
    \put(52.4,15){{$2$}}
    \put(86,17){{$1$}}
    \put(84.3,13.5){{$2$}}
    \end{overpic}
    \caption{Gust encounter $(\phi,z_0)=(3\pi/2,0.60)$: the streamwise vorticity $\omega_x$ contours on transverse planes at $x=0.2$, $x=0.5$, $x=0.8$, $x=1.1$, and $x=1.4$, overlaid by the instantaneous $Q=3.5$ isosurface, at (a)~$t=-0.1$, (b)~$t=0.5$, (c)~$t=0.6$, (d)~$t=0.8$, (e)~$t=1.0$, and $(f)~t=1.1$.}
    \label{fig:vort_z_0d6_n}
\end{figure}

In figure \ref{fig:vort_z_0d6_n} we once again use $Q$ isosurfaces and planar slices of vorticity to clarify the process by which the lobes form and evolve. In this figure we label two lobes, $1$ and $2$, but note that lobe $2$ emerges from the underside of the wing. In panels (a) and (b), the lifting of the boundary layer is apparent. By (c), these lifted structures on each surface have been rotated toward a spanwise orientation, and in panels (d) through (f), the lobes are further twisted around the vortex as it passes the trailing edge. During this process, the vorticity in the lobe cores are amplified by the gust's induced strain, due to the favorable alignment of $T_z$ and $\omega_z$ as shown in figure~\ref{fig:stretch}(c).

From the surface pressure signatures of the three gust encounters we analyzed in this section, the vertical vortex gust always imprints a characteristic low-pressure core on the wing surface and edges, and, depending on favorable coordination of its induced flow with the free stream, can move the stagnation region from the apex to one of the leading edges. These obvious surface features have enabled us to explain many of the observed behaviors in the drag, side force, and yaw moment, and in some cases, aspects of the lift and other moments. However, there still remain unexplained behaviors in these latter components that are not easily apparent in the surface pressure. By connecting topological features in the skin friction lines with the flow kinematics, we have illuminated the emergence of lobes and their subsequent elongation and twisting around the gust core. However, we have not yet clarified the role that these lobes play in disturbing the force and moment on the wing, since their own signatures in the surface traction are obscured by the core itself. Thus, in the next section, we will use force element analysis to help us understand the lobes' contribution to lift.



\subsubsection{Force element analysis}\label{subsec:force_element}

After revealing the formation and evolution of the lobed vortical structures in the three representative gust encounters, let us analyze how these lobes are responsible to the aerodynamic lift variation. To answer that, we utilize the force element theory developed by Chang et al.~\cite{Chang1992,Lee2012}, which has been applied to reveal the contributions of flow structures in the aerodynamic forces in various flow problems \cite{odaka_square_2026,Zhang2020,Chiu2023}. In this analysis technique, an auxiliary potential $\phi_i$ is identified by solving the Laplace equation $\boldsymbol{\nabla}^2 \phi_i=0$ with boundary condition $-\boldsymbol n \cdot \boldsymbol \nabla \phi_i=\boldsymbol n \cdot \boldsymbol e_i$, where the subscript $i=\{ x,y,z \}$ represents the spatial direction $x$, $y$, and $z$, respectively, and the $\boldsymbol{e}_i$ represents the unit vector in the corresponding direction. By taking the inner product of the Navier-Stokes equations with the potential velocity $\boldsymbol \nabla \phi_i$, the aerodynamic force can be expressed as
\begin{equation}
\label{eq:forceelement}
    F_i = \int_V (\boldsymbol{\omega} \times \boldsymbol{u}) \cdot \boldsymbol{\nabla} \phi_i dV + \frac{1}{Re} \int_S (\boldsymbol{\omega} \times \boldsymbol{n}) \cdot (\boldsymbol{\nabla} \phi_i + \boldsymbol{e_i}) dS,
\end{equation}
where the first integral is taken over the volume of fluid and the second over the surface of the wing. The volume integrand $(\boldsymbol{\omega} \times \boldsymbol{u}) \cdot \boldsymbol{\nabla} \phi_i$, denoted as $f_i$ in the following discussion, is referred to as the force element, and represents the contribution of fluid vorticity to the overall component of force $F_i$. By the natural geometric decay of the auxiliary potential, this contribution is concentrated in the vicinity of the wing. (Note that a moment element can also be derived using the same idea.) In the current study, when we consider the $y$ direction, the volume integral contributes around $75\%$ in the total lift. However, it only comprises $20\%$ in the total drag when we consider the $x$ direction, and only around $50\%$ in the total side force for the $z$ direction. For this reason, we only discuss the lift element $f_y=\boldsymbol (\boldsymbol{\omega} \times \boldsymbol{u})\cdot \boldsymbol \nabla \phi_y$ here.

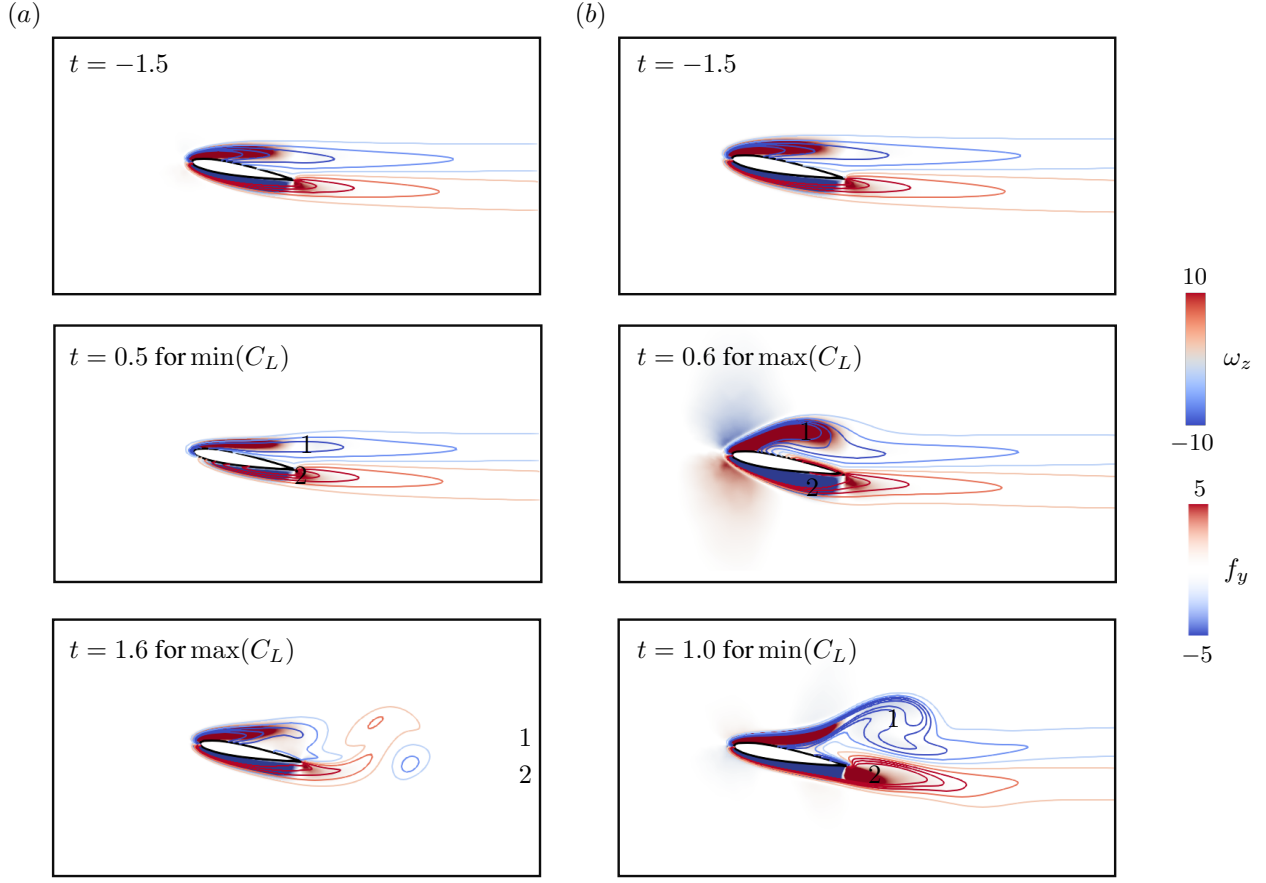
\begin{figure}[htbp]
    \centering
    \begin{overpic}[width=1.0\linewidth]{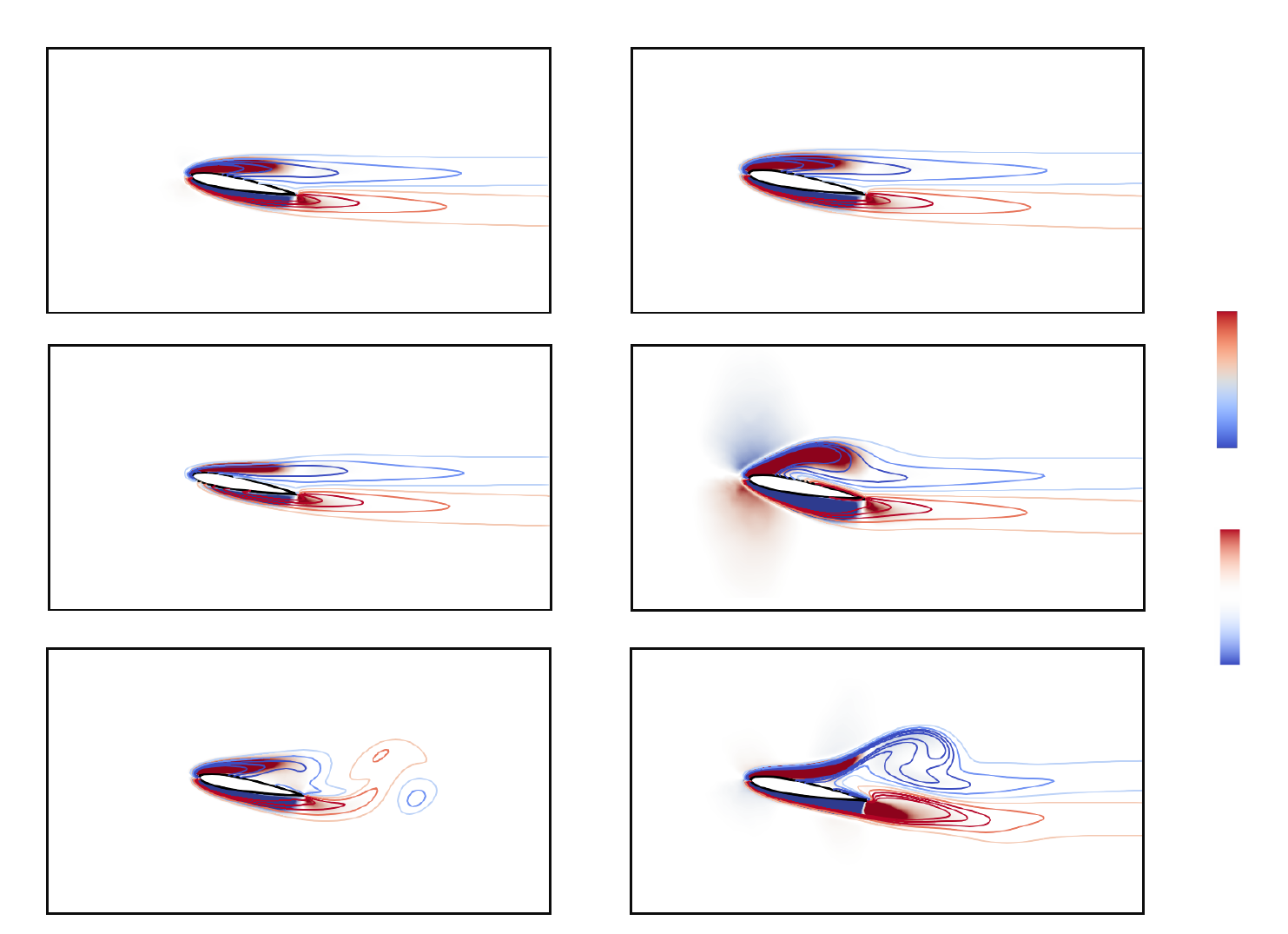}
    \put(97.5,25.5){{$f_y$}}
    \put(94.1,18.7){\small{$-5$}}
    \put(95.1,32){\small{$5$}}
    \put(97.5,42.5){{$\omega_z$}}
    \put(93.2,35.7){\small{$-10$}}
    \put(94.2,49){\small{$10$}}
    \put(0,70){{$(a)$}}
    \put(45.5,70){{$(b)$}}
    \put(5,66){{$t=-1.5$}}
    \put(5,42.5){{$t=0.5~ \text{for} \min(C_L)$}}
    \put(5,19){{$t=1.6~ \text{for} \max(C_L)$}}
    \put(50.5,66){{$t=-1.5$}}
    \put(50.5,42.5){{$t=0.6~ \text{for} \max(C_L)$}}
    \put(50.5,19){{$t=1.0~ \text{for} \min(C_L)$}}
    \put(23.5,35.5){{$1$}}
    \put(23,33){{$2$}}
    \put(41,12){{$1$}}
    \put(41,9){{$2$}}
    \put(63.5,36.5){{$1$}}
    \put(64,32){{$2$}}
    \put(70.5,13.5){{$1$}}
    \put(69,9){{$2$}}
    \end{overpic}
    \caption{The lift element $f_y$ (filled contours) at $z=0.35$ slice, overlaid by the spanwise vorticity $\omega_z$ (line contours), for gust encounters (a) $(\phi,z_0)=(\pi/2,0.60)$ and (b) $(\phi,z_0)=(3\pi/2,0.60)$.}
    \label{fig:vortz_fy}
\end{figure}

Let us first see the distribution of $f_y$ for the two tip-centered gust encounters. The lift elements in these encounters are adequately visualized in a single streamwise plane ($z=0.35$) along with contours of the spanwise vorticity in figure \ref{fig:vortz_fy}. In the top row of panels at $t=-1.5$, the gust is sufficiently far from the wing to have an influence on the wing. We observe that the boundary layer on the suction side (which is marginally separated at this spanwise station) and the free shear layer from the trailing edge in the pressure side both contribute positively to the lift generation, whereas the boundary layer on the pressure side has a negative contribution to the lift. Similar observations have been made for an impulsively-start flat plate wing \cite{Lee2012} and a NACA $0015$ rectangular wing \cite{Zhang2020}.

In the two remaining rows of panels, we show the distributions of $f_y$ at instants of minimum or maximum lift during the respective gust encounters. For the counter-clockwise gust encounter, in column (a), we can see that the boundary layers on both sides of the wing are thinner at $t=0.5$ and the spanwise vorticity weaker due to the formation of the lobes, as we discussed in section \ref{subsec:surface_signatures}. The corresponding lift contributions of both boundary layers is thereby weakened. However, since lobe $1$ is more developed than $2$ at this instant, the lift contribution from the upper side gets weakened more significantly than from the lower, resulting in the minimum lift. At $t=1.6$, when the gust has convected into the wake and the boundary layers are recovering to the baseline flow, the lift element distribution has nearly recovered to its baseline, as well. However, it remains slightly elevated on the upper surface, and thus the lift is at its maximum and slightly larger than at steady state.

In the two lower panels in column (b), we show the clockwise tip-centered gust encounter. We first note, at $t=0.6$, that the two diffuse regions of lift elements just upstream of the leading edge represent the self-canceling contributions from the gust itself. More important to maximum lift at this instant are the lobes, and in particular, the concentrated core of lobe $1$ shows a significant contribution, while the negative contribution from lobe $2$, though also elevated, is not as strong. At $t=1.0$ when the lift is minimum, the gust is just passing the trailing edge. Both lobe $2$ and the shear layer associated with lobe $1$ have notable contributions to the lift generation, while lobe $1$ itself shows a moderate contribution to lift reduction. The additional contributions these features make beyond the baseline nearly negate each other, and thus the minimum lift is only slightly lower than steady state. 

\begin{figure}[htbp]
    \centering
    \begin{overpic}[width=1.0\linewidth]{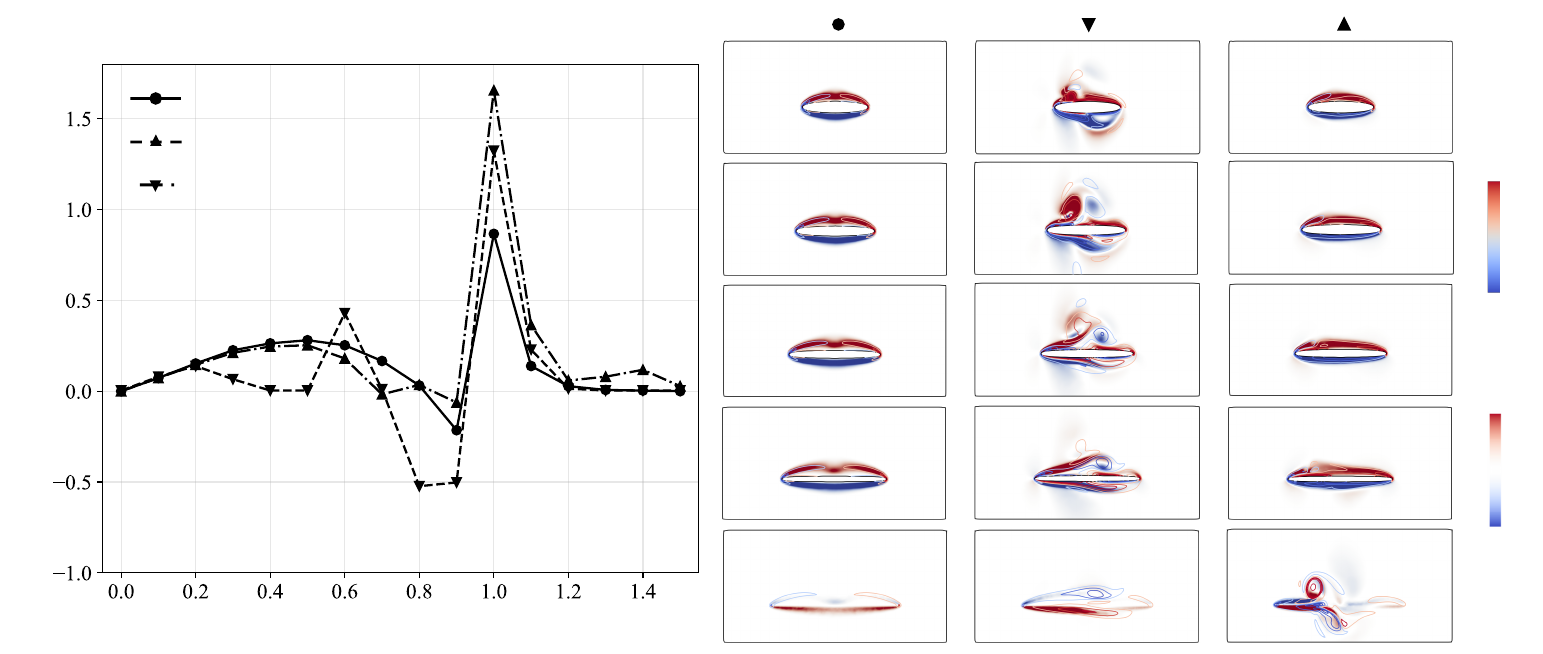}
    \put(25,1.5){{$x$}}
    \put(1,19){\rotatebox{90}{$C_{L_{v,x}}$}}
    \put(12.5,35.5){\small{baseline}}
    \put(12.5,32.8){\small{$\max(C_L)$}}
    \put(12.5,30){\small{$\min(C_L)$}}
    \put(47.5,38){\small{$x=0.5$}}
    \put(47.5,30.1){\small{$x=0.6$}}
    \put(47.5,22.2){\small{$x=0.7$}}
    \put(47.5,14.3){\small{$x=0.8$}}
    \put(47.5,6.3){\small{$x=1.0$}}
    \put(98,11.6){\small{$f_y$}}
    \put(95.5,6.8){\tiny{$-5$}}
    \put(96.5,16.3){\tiny{$5$}}
    \put(98,26.6){\small{$\omega_x$}}
    \put(95,22){\tiny{$-10$}}
    \put(95.9,31.3){\tiny{$10$}}
    \put(69,36){\tiny{$3$}}
    \put(69,28.9){\tiny{$3$}}
    \put(69.5,21){\tiny{$3$}}
    \put(71.1,20){\tiny{$1$}}
    \put(71.3,12){\tiny{$1$}}
    \put(71.6,10.2){\tiny{$2$}}
    \put(70.7,3.5){\tiny{$1$}}
    \put(70.2,2.2){\tiny{$2$}}
    \put(84.7,4.1){\tiny{$3$}}
    \end{overpic}
    \caption{The area integrals $C_{L_{v,x}}$ and contours of the lift element $f_y$ for the gust encounter $(\phi,z_0)=(\pi/2,0)$. Left panel: area integral of lift elements over transverse planes versus $x$; right panel: lift element $f_y$ (filled contours) at five selected $x$-constant slices, overlaid by the streamwise vorticity $\omega_x$ (line contours). The $\min(C_L)$ corresponds to $t=0.7$ and the $\max(C_L)$ corresponds to $t=1.2$.}
    \label{fig:force_ele_z_0}
\end{figure}
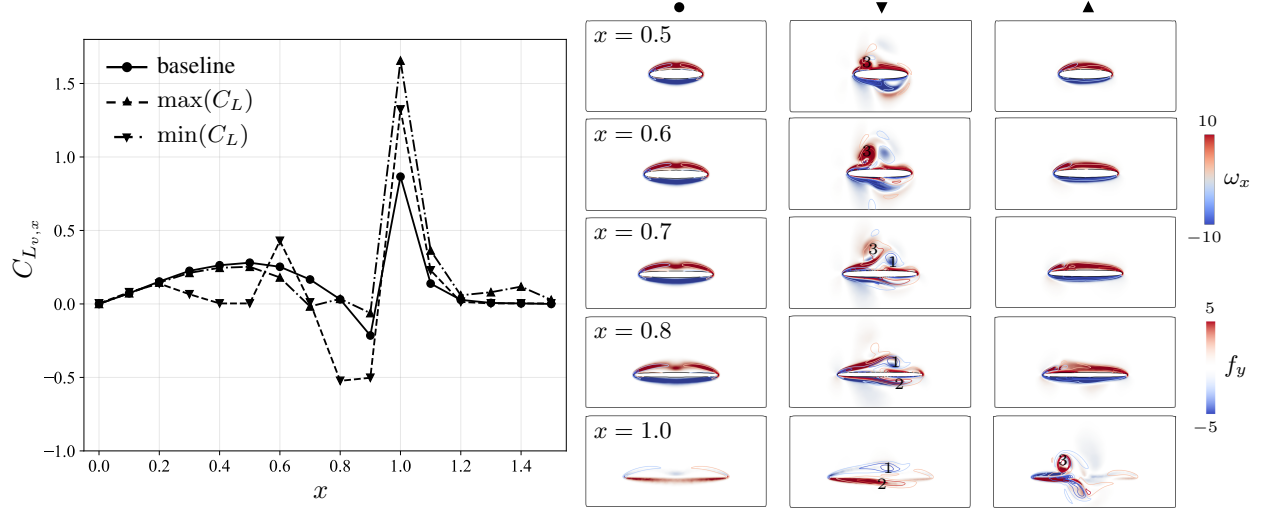

The lift element analysis of the root-centered vortex gust encounter is more complex. For better understanding of the lift element distribution in the three-dimensional space, in figure~\ref{fig:force_ele_z_0}, we inspect the contours of $f_y$ in selected transverse ($x$) planes, and compute the area integral $C_{L_{v,x}}$ for the lift element $f_y$ over each slice as
\begin{equation}
    C_{L_{v,x}} = \frac{\int_{S_x} f_y dS}{\frac{1}{2}\rho U_\infty^2A_w},
\end{equation}
where $S_x$ is a particular transverse plane. The sum of these area integrals over the selected planes approximates the vortex contribution to lift in equation \eqref{eq:forceelement}.

In the baseline, $C_{L_{v,x}}$ first increases up to $x=0.5$, and then decreases and becomes slightly negative at $x=0.9$. The area integral increases rapidly to a peak in the wake just beyond the trailing edge at $x=1.0$, and then decays to zero in the more distant wake. The decrease in the aft portion and the small negative value at $x=0.9$ are due to the subtle balance between the boundary layers on each side, to which the nearly separated boundary layer on the upper side makes a decreasing and ultimately smaller contribution. In contrast, the strong positive peak at $x=1.0$ is attributable to the free shear layer emerging from the lower side of the wing, as we showed in the baseline panels in figure \ref{fig:vortz_fy} and visible in the transverse panel on the right side of figure~\ref{fig:force_ele_z_0}. This contribution is unmatched by any flow feature on the upper side. 

To explain the maximum and minimum lift during this encounter, we will focus on disruptions to these baseline features. At the instant of minimum lift ($t = 0.7$), the streamwise $C_{L_{v,x}}$ profile indicates that the overall lift reduction is attributable to strong suppression of $C_{L_{v,x}}$ near the trailing edge between $x = 0.7$ and $0.9$. By inspecting the contours of $f_y$ in planes $x=0.7$ and $0.8$, we observe that lobes $1$ and $2$ both make negative contributions, and furthermore, lobe $1$ displaces the baseline's positive contribution from the upper side boundary layer. It is notable, however, that this decrease in lift is mitigated by a strong contribution from lobe $3$, apparent in the transverse plane at $x=0.6$, and in the near wake at $x = 1.0$, where the contribution from lobe 2 becomes positive.

At the instant of maximum lift ($t=1.2$), the profile of $C_{L_{v,x}}$ shows that the contribution from the near wake is significantly increased, whereas the contribution from the recovering boundary layers (i.e., the flow between $x=0.0$ and $0.9$) remains slightly decreased. From the contours of the $f_y$ at this time instant, we see that the increased lift contribution from the wake is attributable to the positive contribution of lobe $3$, as shown in $x=1.0$ slice. 

\subsection{Effect of gust size \texorpdfstring{$D$}{D} and strength \texorpdfstring{$G$}{G} on the encounters}\label{subsec:DandG}

In the previous sections, we found that each encounter with a vertical vortex gust is characterized by common features---a low-pressure core imprinted on the surface, lifting of the boundary layers, and the subsequent formation and straining of lobed structures---the balance of these features is strongly affected by the lateral position of the vortex encounter, and thus the force and moment responses exhibit distinct features in each case rather than simple trends.

\begin{figure}[htbp]
    \centering
    \begin{overpic}[width=1.0\linewidth]{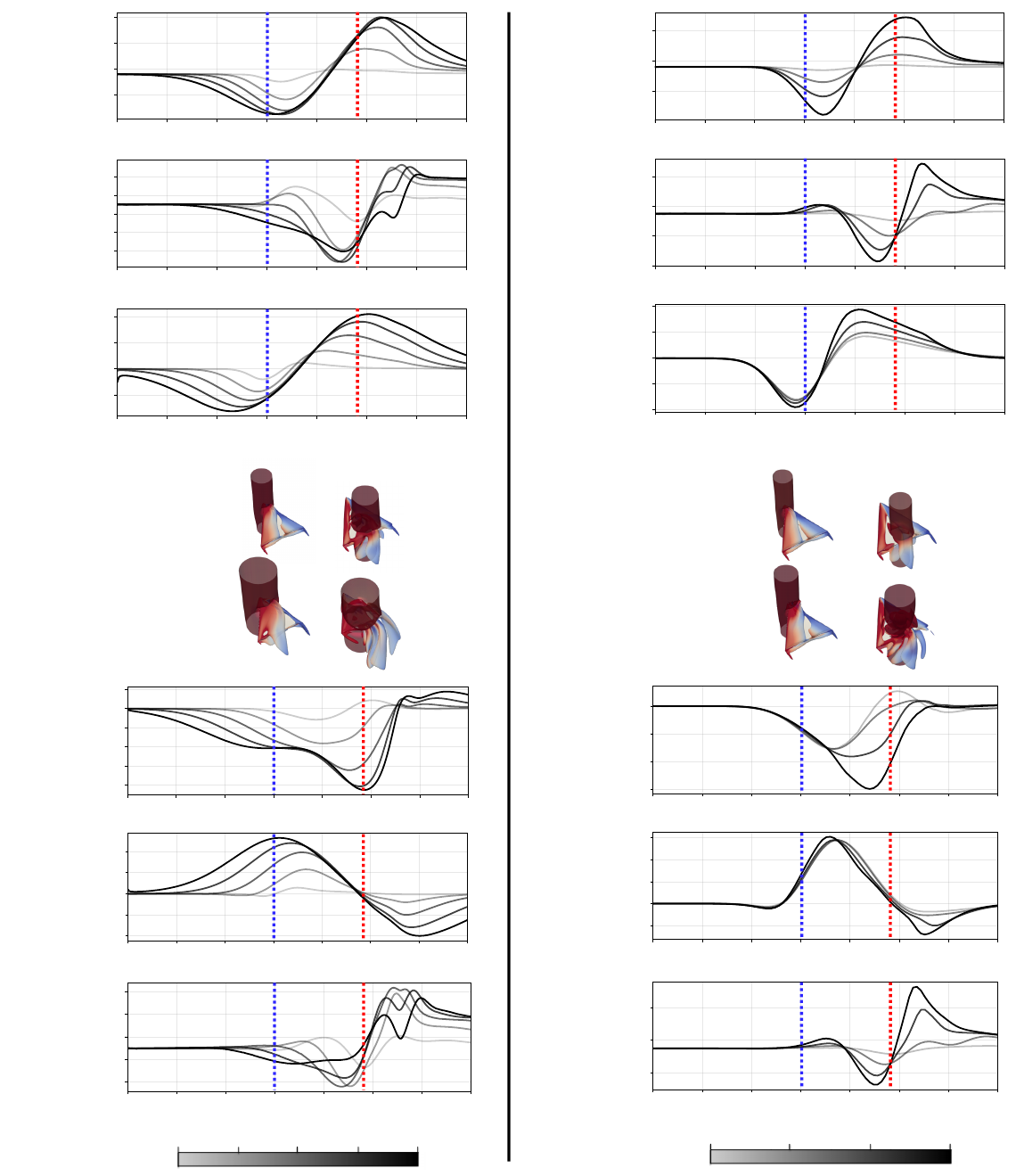}
    \put(21.9,100){\color{blue}{\Large$t_1$}}
    \put(29.5,100){\color{red}{\Large$t_2$}}
    \put(12,54.9){{$D=0.5$}}
    \put(12,46.9){{$D=1.0$}}
    \put(67.5,100){\color{blue}{\Large$t_1$}}
    \put(75,100){\color{red}{\Large$t_2$}}
    \put(56.6,54.9){{$G=0.5$}}
    \put(56.6,46.9){{$G=1.5$}}
    \put(7.9,62.8){\small{$-1.5$}}
    \put(12,62.8){\small{$-1.0$}}
    \put(16,62.8){\small{$-0.5$}}
    \put(21.6,62.8){\small{$0.0$}}
    \put(25.7,62.8){\small{$0.5$}}
    \put(30,62.8){\small{$1.0$}}
    \put(34.1,62.8){\small{$1.5$}}
    \put(38.3,62.8){\small{$2.0$}}

    \put(7.9,75.4){\small{$-1.5$}}
    \put(12,75.4){\small{$-1.0$}}
    \put(16,75.4){\small{$-0.5$}}
    \put(21.6,75.4){\small{$0.0$}}
    \put(25.7,75.4){\small{$0.5$}}
    \put(30,75.4){\small{$1.0$}}
    \put(34.1,75.4){\small{$1.5$}}
    \put(38.3,75.4){\small{$2.0$}}

    \put(7.9,87.9){\small{$-1.5$}}
    \put(12,87.9){\small{$-1.0$}}
    \put(16,87.9){\small{$-0.5$}}
    \put(21.6,87.9){\small{$0.0$}}
    \put(25.7,87.9){\small{$0.5$}}
    \put(30,87.9){\small{$1.0$}}
    \put(34.1,87.9){\small{$1.5$}}
    \put(38.3,87.9){\small{$2.0$}}

    \put(8.3,5.5){\small{$-1.5$}}
    \put(12.3,5.5){\small{$-1.0$}}
    \put(16.7,5.5){\small{$-0.5$}}
    \put(22.1,5.5){\small{$0.0$}}
    \put(26.3,5.5){\small{$0.5$}}
    \put(30.4,5.5){\small{$1.0$}}
    \put(34.5,5.5){\small{$1.5$}}
    \put(38.7,5.5){\small{$2.0$}}

    \put(8.3,18.2){\small{$-1.5$}}
    \put(12.3,18.2){\small{$-1.0$}}
    \put(16.7,18.2){\small{$-0.5$}}
    \put(22.1,18.2){\small{$0.0$}}
    \put(26.3,18.2){\small{$0.5$}}
    \put(30.4,18.2){\small{$1.0$}}
    \put(34.5,18.2){\small{$1.5$}}
    \put(38.7,18.2){\small{$2.0$}}

    \put(8.3,30.7){\small{$-1.5$}}
    \put(12.3,30.7){\small{$-1.0$}}
    \put(16.7,30.7){\small{$-0.5$}}
    \put(22.1,30.7){\small{$0.0$}}
    \put(26.3,30.7){\small{$0.5$}}
    \put(30.4,30.7){\small{$1.0$}}
    \put(34.5,30.7){\small{$1.5$}}
    \put(38.7,30.7){\small{$2.0$}}

    \put(53.3,5.7){\small{$-1.5$}}
    \put(57.3,5.7){\small{$-1.0$}}
    \put(61.7,5.7){\small{$-0.5$}}
    \put(67.1,5.7){\small{$0.0$}}
    \put(71.3,5.7){\small{$0.5$}}
    \put(75.4,5.7){\small{$1.0$}}
    \put(79.5,5.7){\small{$1.5$}}
    \put(83.7,5.7){\small{$2.0$}}

    \put(53.3,17.9){\small{$-1.5$}}
    \put(57.3,17.9){\small{$-1.0$}}
    \put(61.7,17.9){\small{$-0.5$}}
    \put(67.1,17.9){\small{$0.0$}}
    \put(71.3,17.9){\small{$0.5$}}
    \put(75.4,17.9){\small{$1.0$}}
    \put(79.5,17.9){\small{$1.5$}}
    \put(83.7,17.9){\small{$2.0$}}

    \put(53.3,30.7){\small{$-1.5$}}
    \put(57.3,30.7){\small{$-1.0$}}
    \put(61.7,30.7){\small{$-0.5$}}
    \put(67.1,30.7){\small{$0.0$}}
    \put(71.3,30.7){\small{$0.5$}}
    \put(75.4,30.7){\small{$1.0$}}
    \put(79.5,30.7){\small{$1.5$}}
    \put(83.7,30.7){\small{$2.0$}}

    \put(53.3,62.8){\small{$-1.5$}}
    \put(57.3,62.8){\small{$-1.0$}}
    \put(61.7,62.8){\small{$-0.5$}}
    \put(67.1,62.8){\small{$0.0$}}
    \put(71.3,62.8){\small{$0.5$}}
    \put(75.5,62.8){\small{$1.0$}}
    \put(79.5,62.8){\small{$1.5$}}
    \put(83.7,62.8){\small{$2.0$}}

    \put(53.3,75.4){\small{$-1.5$}}
    \put(57.3,75.4){\small{$-1.0$}}
    \put(61.7,75.4){\small{$-0.5$}}
    \put(67.1,75.4){\small{$0.0$}}
    \put(71.3,75.4){\small{$0.5$}}
    \put(75.5,75.4){\small{$1.0$}}
    \put(79.5,75.4){\small{$1.5$}}
    \put(83.7,75.4){\small{$2.0$}}

    \put(53.3,87.9){\small{$-1.5$}}
    \put(57.3,87.9){\small{$-1.0$}}
    \put(61.7,87.9){\small{$-0.5$}}
    \put(67.1,87.9){\small{$0.0$}}
    \put(71.3,87.9){\small{$0.5$}}
    \put(75.5,87.9){\small{$1.0$}}
    \put(79.5,87.9){\small{$1.5$}}
    \put(83.7,87.9){\small{$2.0$}}

    \put(6,65.6){\small{$-0.1$}}
    \put(7.3,67.9){\small{$0.0$}}
    \put(7.3,70.1){\small{$0.1$}}
    \put(7.3,72.3){\small{$0.2$}}
    \put(7.3,67.9){\small{$0.0$}}
    \put(3,68){\rotatebox{90}{\large$C_S$}}

    \put(6.5,78){\small{$0.15$}}
    \put(6.5,79.6){\small{$0.20$}}
    \put(6.5,81.2){\small{$0.25$}}
    \put(6.5,82.8){\small{$0.30$}}
    \put(6.5,84.4){\small{$0.35$}}
    \put(3,79.9){\rotatebox{90}{\large$C_L$}}

    \put(7.3,91.2){\small{$0.1$}}
    \put(7.3,93.4){\small{$0.2$}}
    \put(7.3,95.6){\small{$0.3$}}
    \put(7.3,97.8){\small{$0.4$}}
    \put(3,92.7){\rotatebox{90}{\large$C_D$}}

    \put(6.3,7.3){\small{$0.050$}}
    \put(6.3,9.3){\small{$0.075$}}
    \put(6.3,11.3){\small{$0.100$}}
    \put(6.3,13.2){\small{$0.125$}}
    \put(6.3,15.1){\small{$0.150$}}
    \put(3,9.4){\rotatebox{90}{\large$C_{M_p}$}}

    \put(5,19.8){\small{$-0.050$}}
    \put(5,21.5){\small{$-0.025$}}
    \put(6.3,23.3){\small{$0.000$}}
    \put(6.3,25.1){\small{$0.025$}}
    \put(6.3,26.9){\small{$0.050$}}
    \put(3,22.7){\rotatebox{90}{\large$C_{M_y}$}}

    \put(5.8,32.6){\small{$-0.08$}}
    \put(5.8,34.2){\small{$-0.06$}}
    \put(5.8,35.8){\small{$-0.04$}}
    \put(5.8,37.4){\small{$-0.02$}}
    \put(7.2,39){\small{$0.00$}}
    \put(7.2,40.6){\small{$0.02$}}
    \put(3,35.1){\rotatebox{90}{\large$C_{M_r}$}}

    \put(50.8,64.4){\small{$-0.10$}}
    \put(50.8,66.6){\small{$-0.05$}}
    \put(52.1,68.8){\small{$0.00$}}
    \put(52.1,71){\small{$0.05$}}
    \put(52.1,73.2){\small{$0.10$}}
    \put(47.5,68){\rotatebox{90}{\large$\hat{C_S}$}}

    \put(52.8,76.9){\small{$0.1$}}
    \put(52.8,79.3){\small{$0.2$}}
    \put(52.8,81.7){\small{$0.3$}}
    \put(52.8,84.1){\small{$0.4$}}
    \put(47.5,79.9){\rotatebox{90}{\large$\hat{C_L}$}}

    \put(52.8,91.4){\small{$0.1$}}
    \put(52.8,94.1){\small{$0.2$}}
    \put(52.8,96.8){\small{$0.3$}}
    \put(47.5,92.5){\rotatebox{90}{\large$\hat{C_D}$}}

    \put(51.9,8){\small{$0.05$}}
    \put(51.9,10.8){\small{$0.10$}}
    \put(51.9,13.6){\small{$0.15$}}
    \put(46.9,9.4){\rotatebox{90}{\large$\hat{C}_{M_p}$}}

    \put(50.6,20.7){\small{$-0.01$}}
    \put(51.9,22.5){\small{$0.00$}}
    \put(51.9,24.3){\small{$0.01$}}
    \put(51.9,26.1){\small{$0.02$}}
    \put(51.9,27.9){\small{$0.03$}}
    \put(46.9,21.9){\rotatebox{90}{\large$\hat{C}_{M_y}$}}

    \put(50.7,32.4){\small{$-0.06$}}
    \put(50.7,34.6){\small{$-0.04$}}
    \put(50.7,36.9){\small{$-0.02$}}
    \put(52.1,39.3){\small{$0.00$}}
    \put(46.9,34.5){\rotatebox{90}{\large$\hat{C}_{M_r}$}}

    \put(25,4.4){\large{$t$}}
    \put(70,4.4){\large{$t$}}
    
    \put(24.5,-0.7){\small{$D$}}
    \put(13.5,2.7){\small{$0.25$}}
    \put(18.5,2.7){\small{$0.50$}}
    \put(23.5,2.7){\small{$0.75$}}
    \put(28.5,2.7){\small{$1.00$}}
    \put(33.5,2.7){\small{$1.25$}}
    
    \put(70,-0.7){\small{$G$}}
    \put(59.2,2.75){\small{$0.1$}}
    \put(65.9,2.75){\small{$0.5$}}
    \put(72.8,2.75){\small{$1.0$}}
    \put(79.5,2.75){\small{$1.5$}}

    \put(0,0){\tikz[overlay] \draw[ultra thick, blue, dotted] (4.37,11.85) -- (4.37,11.49);}
    \put(0,0){\tikz[overlay] \draw[ultra thick, red, dotted] (5.79,11.85) -- (5.79,11.49);}

    \put(0,0){\tikz[overlay] \draw[ultra thick, blue, dotted] (13.1,11.85) -- (13.1,11.49);}
    \put(0,0){\tikz[overlay] \draw[ultra thick, red, dotted] (14.52,11.85) -- (14.52,11.49);}

    \put(0,0){\tikz[overlay] \draw[ultra thick, blue, dotted] (4.45,7.97) -- (4.45,8.32);}
    \put(0,0){\tikz[overlay] \draw[ultra thick, red, dotted] (5.89,7.97) -- (5.89,8.32);}

    \put(0,0){\tikz[overlay] \draw[ultra thick, blue, dotted] (13.03,7.97) -- (13.03,8.32);}
    \put(0,0){\tikz[overlay] \draw[ultra thick, red, dotted] (14.48,7.97) -- (14.48,8.32);}
    \end{overpic}
    \caption{The aerodynamic responses and instantaneous flow structures during the parametric study of gust size $D$ (left panel) and gust strength $G$ (right panel).}
    \label{fig:aero_D_G}
\end{figure}
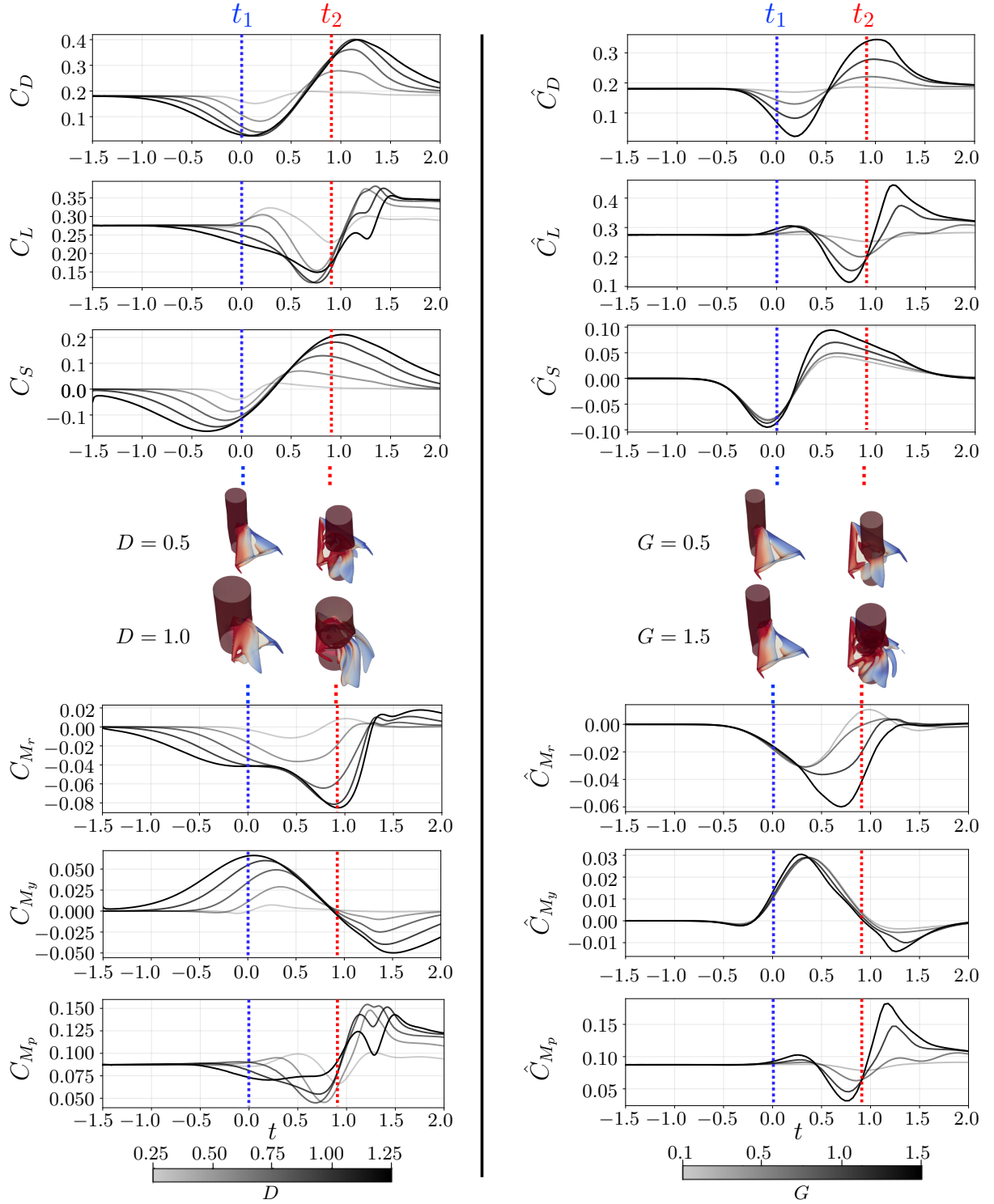

The next important question is to ask is whether the aerodynamic response will also vary in distinct fashion when the gust's size $D$ or its strength $G$ is varied. To answer that, we perform a sweep over these parameters, shown in figure \ref{fig:aero_D_G}. In the case of strength $G$, we are interested in the degree of nonlinear dependence on this parameter. Here, we present re-scaled values of the force and moment components,
\begin{equation}
    \hat{C_i}(t) = C_{i}^0 + \frac{C_i(t)-C_{i}^0}{G}, \qquad \hat{C}_{M_i}(t) = C_{M_{i}}^0 + \frac{C_{M_i}(t)-C_{M_{i}}^0}{G}. 
\end{equation}
where the subscript index $i$ represents each component of force coefficients and moment coefficients, and the superscript $0$ represents the baseline value.

It is clear that, in contrast to the distinct variation of the aerodynamic response with lateral position, both the parameters here lead to largely monotonic variation of the aerodynamic response. In the study of gust size $D$, the $Q$ isosurfaces at $t_1=0.0$ and $t_2=0.9$ show that similar flow structures are formed in all cases, though the lobes are induced earlier and have stronger deformations for larger gust size. These larger gust cases give rise to the only feature of note in these studies, in the lift and pitching moment at around $t = 1.25$. The common features in these cases are also clearly reflected in the drag force, side force, and the yaw moment profiles: a larger gust elicits the same basic response, but stronger and more sustained (disrupted earlier and recovered later). 

Similar observations can be made for the study of gust strength $G$. From the instantaneous $Q$ isosurfaces at $t_1=0.0$ when the gust just hits the wing, a similar disturbance to the boundary layer is observed, while the stronger gust produces a stronger disturbance. At $t_2=0.9$ when the gust almost passes the trailing edge, similar primary flow structures are formed, while the stronger gust seems to induce some secondary structures. It is notable that the yaw moment and, to a lesser extent, the side force appear to depend linearly on the gust strength. 

\section{Conclusions}\label{sec:conclusions}

We have numerically investigated the interaction between a tailless NACA $0012$ delta wing with a sweep angle of $60^\circ$, an angle of attack of $10^\circ$ and a Reynolds number of $Re = 1000$, and vertical vortex gusts modeled as a Taylor vortex. Such gust encounters are found to impose strong transient loads on all six degrees of freedom. By systematically varying the gust's lateral position, size, and strength, we have shown that the vertical vortex gust encounters are governed by a common characteristic physical mechanism, and have further revealed how the variation of individual gust parameters influences the interaction and the associated aerodynamic response.

In particular, we have shown that the vertical vortex gust consistently imprints a characteristic low-pressure core on the wing surface and regions of suction and stagnation on the wing's leading edge, which principally determine the responses of drag, side force, and yaw moment. The core's imprint creates regions of adverse pressure gradient and evolving lines of separation. The separation lifts the boundary layers, which then form lobed vortical structures. The produced lobes, which are elongated and twisted by the strain field of the gust, are shown to play an important role in shaping the aerodynamic lift. Depending on the relative positions of the lobes to the wing, their contribution to the lift can be either positive or negative.

This physical mechanism is observable for all the considered gust parameters. In particular, as the gust encounter is laterally shifted from root to tip, the flow features are marginalized to the leading edge nearest the gust and the balance of their contributions to the forces and moments is modified, resulting in distinct variations of the aerodynamic response with lateral position. In contrast, the gust size and strength both elicit largely monotonic variation of the aerodynamic response and do not  modify the balance of the primary flow structures. Specifically, a larger gust results in a stronger and more sustained response. Unsurprisingly, a stronger gust is found to induce a stronger response, but we have also found that some components (side force and yaw moment) are nearly linearly dependent on this strength.

Overall, this study has provided a mechanistic foundation for the interaction between a vertical vortex gust and a tailless delta wing, complementing previous studies that have focused on streamwise and spanwise vortex gust encounters. The findings in this study advance our more general understanding of flow responses and the associated unsteady aerodynamic loads for a tailless delta wing in more complex atmospheric environments. Furthermore, it should provide guidance for future studies to explore control strategies for attenuating the unsteady aerodynamic loads of a delta wing in these complex environments.

\section*{Acknowledgments}
We dedicate this paper to our close friend and colleague, Prof.~David R.~Williams, whose extensive work in aerodynamics has inspired this study and who provided valuable counsel on its early stages. The authors gratefully acknowledge the support provided by the U.S. Air Force Office of Scientific Research (FA9550-26-1-0006) and the U.S. Department of Defense Vannevar Bush Faculty Fellowship (N00014-22-1-2798). This work used the Delta system at the National Center for Supercomputing Applications, which is supported by the National Science Foundation through allocation PHY250240. The authors also thank Anya R. M. Jones and Aimee S. Morgans for enlightening discussions.

\appendix
\section{Grid convergence}\label{app:appendix}

\begin{figure}[htbp]
    \centering
    \begin{overpic}[width=1.0\linewidth]{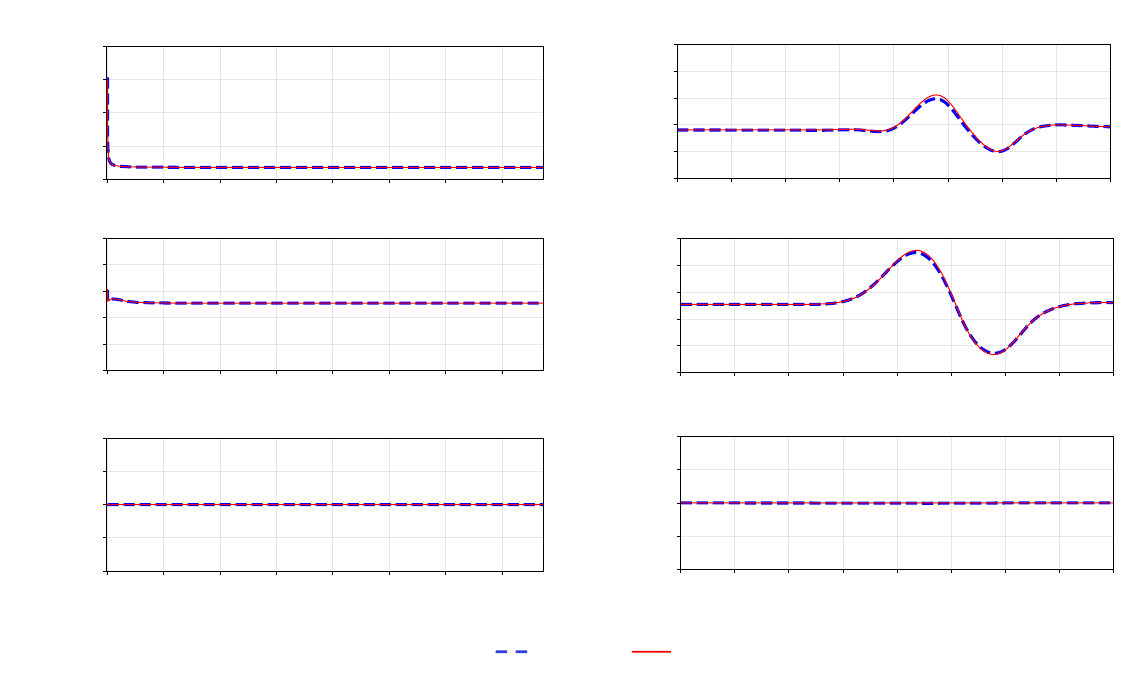}
    \put(1.5,48){\rotatebox{90}{$C_D$}}
    \put(1.5,31.2){\rotatebox{90}{$C_L$}}
    \put(1.5,13.4){\rotatebox{90}{$C_S$}}
    \put(52,48){\rotatebox{90}{$C_D$}}
    \put(52,31.2){\rotatebox{90}{$C_L$}}
    \put(52,13.4){\rotatebox{90}{$C_S$}}
    \put(27,5){{$t$}}
    \put(78,5){{$t$}}
    \put(0,56){{$(a)$}}
    \put(50.5,56){{$(b)$}}

    \put(8.9,7.2){\small{$0$}}
    \put(13.9,7.2){\small{$5$}}
    \put(18.5,7.2){\small{$10$}}
    \put(23.4,7.2){\small{$15$}}
    \put(28.3,7.2){\small{$20$}}
    \put(33.2,7.2){\small{$25$}}
    \put(38.1,7.2){\small{$30$}}
    \put(43,7.2){\small{$35$}}

    \put(8.9,24.7){\small{$0$}}
    \put(13.9,24.7){\small{$5$}}
    \put(18.5,24.7){\small{$10$}}
    \put(23.4,24.7){\small{$15$}}
    \put(28.3,24.7){\small{$20$}}
    \put(33.2,24.7){\small{$25$}}
    \put(38.1,24.7){\small{$30$}}
    \put(43,24.7){\small{$35$}}

    \put(8.9,41.4){\small{$0$}}
    \put(13.9,41.4){\small{$5$}}
    \put(18.5,41.4){\small{$10$}}
    \put(23.4,41.4){\small{$15$}}
    \put(28.3,41.4){\small{$20$}}
    \put(33.2,41.4){\small{$25$}}
    \put(38.1,41.4){\small{$30$}}
    \put(43,41.4){\small{$35$}}

    \put(57,7.2){\small{$-2.0$}}
    \put(62,7.2){\small{$-1.5$}}
    \put(66.8,7.2){\small{$-1.0$}}
    \put(71.7,7.2){\small{$-0.5$}}
    \put(77.5,7.2){\small{$0.0$}}
    \put(82.3,7.2){\small{$0.5$}}
    \put(87,7.2){\small{$1.0$}}
    \put(91.6,7.2){\small{$1.5$}}
    \put(96.3,7.2){\small{$2.0$}}

    \put(57,24.7){\small{$-2.0$}}
    \put(62,24.7){\small{$-1.5$}}
    \put(66.8,24.7){\small{$-1.0$}}
    \put(71.7,24.7){\small{$-0.5$}}
    \put(77.5,24.7){\small{$0.0$}}
    \put(82.3,24.7){\small{$0.5$}}
    \put(87,24.7){\small{$1.0$}}
    \put(91.6,24.7){\small{$1.5$}}
    \put(96.3,24.7){\small{$2.0$}}

    \put(57,41.4){\small{$-2.0$}}
    \put(62,41.4){\small{$-1.5$}}
    \put(66.8,41.4){\small{$-1.0$}}
    \put(71.7,41.4){\small{$-0.5$}}
    \put(77.5,41.4){\small{$0.0$}}
    \put(82.3,41.4){\small{$0.5$}}
    \put(87,41.4){\small{$1.0$}}
    \put(91.6,41.4){\small{$1.5$}}
    \put(96.3,41.4){\small{$2.0$}}

    \put(4,8.9){\small{$-0.2$}}
    \put(4,11.8){\small{$-0.1$}}
    \put(5.5,14.6){\small{$0.0$}}
    \put(5.5,17.5){\small{$0.1$}}
    \put(5.5,20.3){\small{$0.2$}}

    \put(4,26.3){\small{$-1.0$}}
    \put(4,28.6){\small{$-0.5$}}
    \put(5.5,30.9){\small{$0.0$}}
    \put(5.5,33.2){\small{$0.5$}}
    \put(5.5,35.5){\small{$1.0$}}
    \put(5.5,37.8){\small{$1.5$}}

    \put(5.5,43.1){\small{$0.0$}}
    \put(5.5,45.9){\small{$0.5$}}
    \put(5.5,48.8){\small{$1.0$}}
    \put(5.5,51.7){\small{$1.5$}}
    \put(5.5,54.6){\small{$2.0$}}

    \put(54.2,8.9){\small{$-0.2$}}
    \put(54.2,11.8){\small{$-0.1$}}
    \put(55.7,14.7){\small{$0.0$}}
    \put(55.7,17.6){\small{$0.1$}}
    \put(55.7,20.5){\small{$0.2$}}

    \put(54.2,26.3){\small{$-1.0$}}
    \put(54.2,28.5){\small{$-0.5$}}
    \put(55.7,30.8){\small{$0.0$}}
    \put(55.7,33.1){\small{$0.5$}}
    \put(55.7,35.5){\small{$1.0$}}
    \put(55.7,37.8){\small{$1.5$}}

    \put(55.7,43.2){\small{$0.0$}}
    \put(55.7,45.6){\small{$0.1$}}
    \put(55.7,47.9){\small{$0.2$}}
    \put(55.7,50.2){\small{$0.3$}}
    \put(55.7,52.5){\small{$0.4$}}
    \put(55.7,54.8){\small{$0.5$}}

    \put(47,1.7){\small{$18.1M$}}
    \put(59.5,1.7){\small{$29.5M$}}
    \end{overpic}
    \caption{$(a)$ the aerodynamic force histories before a gust encounter under two grid resolutions; $(b)$ the aerodynamic force histories after a gust encounter under two grid resolutions.}
    \label{fig:convergence}
\end{figure}

To verify the spatial convergence of the numerical results, we examine one gust encounter (case ID 1) with two meshes. The medium mesh has $18.1$ millions cells, where the number of quadrilateral faces on the wing surface is set to $n_w=21372$, and the initial off-wall spacing in viscous wall units is set to $y_0^+=0.21$. We refine the medium mesh in all three directions such that the refined mesh has $29.5$ millions cells with $n_w=28812$ and $y_0^+=0.14$. We show the aerodynamic force results, before and after the introduction of gust, under the two different meshes in figure \ref{fig:convergence}. Before the introduction of the gust, we can observe that the flow reaches steady state after the initial transient. After the introduction of the gust, we define $t=0$ as the time instant when the gust vortex center would hit the apex of the delta wing if convecting only with the free stream. Note that $\boldsymbol x_0$ for the two cases considered here is set to $-2$, so the first time instant is $t=-2$ after the introduction of the gust. Since we find that the aerodynamic force starts to be influenced by the gust only after $t \geq -1$, for the remaining studies we choose $x_0=-1.5$ for saving computational resources. The aerodynamic force profiles under the two meshes agree with each other very well, which indicates that both meshes should be adequate to resolve the gust encounter. However, since the gust parameter space considered in the present study is relatively large, the refined mesh (i.e., $29.5$ million cells) becomes a safer choice and is used throughout the study.

\bibliographystyle{unsrt-init}  
\bibliography{main}

\end{document}